\documentclass{article}

\pdfoutput = 1

\usepackage{arxiv}

\usepackage[utf8]{inputenc} 
\usepackage[T1]{fontenc}    
\usepackage{hyperref}       
\usepackage{url}            
\usepackage{booktabs}       
\usepackage{amsfonts}       
\usepackage{nicefrac}       
\usepackage{microtype}      
\usepackage{lipsum}		
\usepackage{graphicx}
\usepackage[numbers,square]{natbib}
\usepackage{doi}
\usepackage{authblk}
\usepackage{comment}
\usepackage{subcaption}
\usepackage{amsmath}
\usepackage{amssymb}
\usepackage{siunitx}
\usepackage{xfrac}
\usepackage{mathptmx} 

\title{Integrating anatomy and electrophysiology in the healthy human heart: 
Insights from biventricular statistical shape analysis using universal coordinates}

\date{} 					

\author[1,*]{Lore Van Santvliet}
\author[2,3]{Elena Zappon}
\author[2]{Matthias A.F. Gsell}
\author[2,4,5]{Franz Thaler}
\author[6,7]{Maarten Blondeel}
\author[8]{Steven Dymarkowski}
\author[9,10]{Guido Claessen}
\author[6,7]{Rik Willems}
\author[3,5]{Martin Urschler}
\author[6,7]{Bert Vandenberk}
\author[2,3]{Gernot Plank}
\author[1]{Maarten De Vos}

\affil[1]{STADIUS Center for Dynamical Systems, Signal Processing and Data Analytics, Department of Electrical Engineering (ESAT), KU Leuven, Leuven, Belgium \\}
\affil[2]{Division of Medical Physics and Biophysics, Gottfried Schatz Research Center, Medical University of Graz, Graz, Austria}
\affil[3]{BioTechMed-Graz, Graz, Austria}
\affil[4]{Institute of Computer Graphics and Vision, Graz University of Technology, Graz, Austria}
\affil[5]{Institute for Medical Informatics, Statistics and Documentation, Medical University of Graz, Graz, Austria}
\affil[6]{Department of Cardiology, University Hospitals Leuven, Leuven, Belgium}
\affil[7]{Department of Cardiovascular Sciences, KU Leuven, Leuven, Belgium}
\affil[8]{Division of Radiology, University Hospitals Leuven, Leuven, Belgium}
\affil[9]{Division of Cardiology, Hartcentrum, Jessa Ziekenhuis, Hasselt, Belgium}
\affil[10]{Department of Medicine and Life Sciences, University of Hasselt, Hasselt, Belgium}
\affil[*]{\textbf{Corresponding author. Email address: lore.vansantvliet@kuleuven.be}}

\renewcommand{\headeright}{Accepted manuscript}
\renewcommand{\undertitle}{\textit{Accepted manuscript version. The final published article is available at \href{https://doi.org/10.1016/j.compbiomed.2025.110230}{https://doi.org/10.1016/j.compbiomed.2025.110230}}}
\renewcommand{\shorttitle}{ }

\hypersetup{
pdftitle={Integrating anatomy and electrophysiology in the healthy human heart: 
Insights from biventricular statistical shape analysis using universal coordinates},
pdfsubject={q-bio.QM},
pdfauthor={Lore Van Santvliet, Elena Zappon, Matthias A.F. Gsell, Franz Thaler, Maarten Blondeel, Steven Dymarkowski, Guido Claessen, Rik Willems, Martin Urschler, Bert Vandenberk, Gernot Plank, Maarten De Vos},
pdfkeywords={First keyword, Second keyword, More},
}

\begin{document}


\maketitle

\begin{abstract}
	A cardiac digital twin is a virtual replica of a patient-specific heart, mimicking its anatomy and physiology. A crucial step of building a cardiac digital twin is anatomical twinning, where the computational mesh of the digital twin is tailored to the patient-specific cardiac anatomy. In a number of studies, the effect of anatomical variation on clinically relevant functional measurements like electrocardiograms (ECGs) is investigated, using computational simulations. While such a simulation environment provides researchers with a carefully controlled ground truth, the impact of anatomical differences on functional measurements in real-world patients remains understudied. In this study, we develop a biventricular statistical shape model and use it to quantify the effect of biventricular anatomy on ECG-derived and demographic features, providing novel insights for the development of digital twins of cardiac electrophysiology. 
To this end, a dataset comprising high-resolution cardiac CT scans from 271 healthy individuals, including athletes, is utilized. Furthermore, a novel, universal, ventricular coordinate-based method is developed to establish lightweight shape correspondence. The performance of the shape model is rigorously established, focusing on its dimensionality reduction capabilities and the training data requirements. The most important variability in healthy ventricles captured by the model is their size, followed by their elongation. These anatomical factors are found to significantly correlate with ECG-derived and demographic features. Additionally, a comprehensive synthetic cohort is made available, featuring ready-to-use biventricular meshes with fiber structures and anatomical region annotations. These meshes are well-suited for electrophysiological simulations.
\end{abstract}



\section{Introduction}
\label{sec1}

\subsection{Cardiac digital twins}
\label{subsec1}

The heart is a complex and highly variable organ, both in shape and function, constituting an essential part of the human body \citep{shaffer_healthy_2014, xiong_comprehensive_2017}. Computational modeling of the heart is a challenging task, usually addressed by complex mathematical models and numerical methods, combining clinical measurements and partial differential equations (PDEs) \citep{gillette_personalized_2022, jung_integrated_2022}. 

Cardiac models can be adapted to represent the actual anatomy and function of individual patients, leading to the concept of a cardiac digital twin (CDT). CDTs are virtual, subject-specific heart models that closely mirror the structure and function of their physical counterparts \citep{sel_building_2024}. Beyond enhancing fundamental understanding of cardiovascular physiology and pathophysiology \citep{lopez-perez_personalized_2019}, direct clinical applications of CDTs include surgical planning \citep{prakosa_personalized_2018}, prediction of clinical outcomes \citep{ohara_personalized_2022}, and risk assessment \citep{arevalo_arrhythmia_2016, cartoski_computational_2019}. As such, these models have the potential to improve patient outcomes via personalized and preventative medicine. CDTs focus on modeling a single aspect of the cardiovascular system, such as electrophysiology (EP), mechanical function, or hemodynamics, or a combination thereof. 

Regardless of the specific aspect(s) of the cardiovascular system that a model targets, the anatomy of the individual needs to be taken into account. As such, defining a computational mesh representing this anatomy is a common step in the construction of all types of CDTs. This discrete, multi-label mesh is typically derived from cardiac computed tomography (CT) or magnetic resonance (MR) imaging of the subject. It delineates the computational domains for discretizing and numerically solving the PDEs governing the system. The process of creating a mesh that matches subject-specific anatomy is known as anatomical twinning \citep{gillette_framework_2021}. 

On a broader scale, digital twins can be extended from individuals to populations, leading to the concept of virtual cohorts. These virtual cohorts enable in silico clinical trials that offer several advantages over traditional approaches, such as reducing the need for real patients to undergo experimental treatments, minimizing the risk of failed trials and associated financial costs, and optimizing trial design \citep{corral-acero_digital_2020, niederer_creation_2020}. Constructing an accurate anatomical representation of the target population is a crucial first step in the pipeline for developing functional, population-level CDTs.

Statistical shape modeling offers a powerful means to study and represent individual variation within a broader population framework \citep{fonseca_cardiac_2011, cootes_use_1994, bai_bi-ventricular_2015, mauger_right_2019, rodero_linking_2021, nagel_bi-atrial_2021, burns_genetic_2024}. Cardiac statistical shape models (SSMs), which capture population-level anatomical variations, closely relate to patient-level CDTs in the applied methodology, since both rely on anatomical twinning. Moreover, these models allow the creation of representative synthetic populations, enhancing virtual cohort studies and providing an anatomical building block for population-level CDTs.

\subsection{Statistical shape modeling}
\label{subsec3}

Statistical shape modeling enables a mathematically rigorous assessment of shape variability within a population through principal component analysis (PCA) \citep{cootes_use_1994, ambellan_statistical_2019}. In the field of computational cardiology, there are numerous applications for such statistical shape models (SSMs), many of which are directly or indirectly related to the concept of CDTs.

(1) \textbf{Synthetic data generation:} An SSM can be leveraged to generate synthetic heart geometries that cover the shape variability present in the training population. These synthetic geometries blend the characteristics of existing hearts and can be limited to remain within the same statistical distribution as the original data. The use of synthetic data offers two main advantages. First, since the synthetic geometries do not correspond to real individuals, they do not contain direct sensitive health information. As such, they can be made publicly available without privacy concerns. Second, SSMs allow for repeated sampling from the statistical shape distribution approximated by the training data, enabling the generation of a virtually unlimited number of synthetic, yet realistic heart geometries. These two elements directly address data availability challenges in CDT development, where high-quality tomographic imaging is often required for anatomical twinning, but scarce \citep{crozier_image-based_2016, gillette_framework_2021}. By offering a scalable alternative, synthetic data generated from SSMs could help alleviate this limitation. Representative, synthetic geometries obtained from SSMs are also of great use in virtual clinical trials \citep{niederer_creation_2020} and large-scale simulation studies \citep{gillette_medalcare-xl_2023}.

(2) \textbf{Dimensionality reduction and machine learning (ML):} The incorporation of ML and deep learning (DL) into CDT construction pipelines holds great promise for the personalization of CDTs using functional data \citep{gillette_framework_2021, dhamala_embedding_2020, coveney_bayesian_2021}. However, these algorithms typically require large volumes of training data, preferably in a suitable input format. Highly complex tomography imaging or computational mesh formats might not be optimal for certain ML and DL frameworks. Statistical shape analysis, with its dimensionality reduction capabilities, can represent complex heart geometries in a more efficient and possibly more compatible format. As such, SSMs enable and facilitate the integration of DL and ML into CDT technology. Moreover, the ability to generate large amounts of synthetic data through SSMs addresses the data-intensive requirements of ML and DL, further supporting their integration into CDT pipelines.  Additionally, cardiac SSMs support applications in automated segmentation algorithms \citep{heimann_statistical_2009, piazzese_chapter_2017}.

(3) \textbf{Statistical analysis:} A cardiac SSM offers valuable, quantitative insights into heart shape variability, which can in turn be linked to other health-related data, such as electrophysiological measurements and demographic information, through statistical analyses \citep{rodero_linking_2021, nagel_bi-atrial_2021, burns_genetic_2024}. As such, cardiac SSM are valuable tools for advancing our fundamental understanding of cardiac anatomy, notably in relation to other clinically relevant features, such as electrocardiograms (ECGs). 

(4) \textbf{Baseline anatomical framework:} SSMs of healthy cardiac anatomy may offer a robust foundation for developing SSMs related to cardiovascular diseases and spatiotemporal variation in heart anatomy, by modeling static, healthy cardiac anatomy as a reference point. 

\subsection{Related work}
\label{subsec5}

Previous work in cardiac SSM development provides meaningful insights into cardiac anatomy of various scopes (biventricular anatomy in \citep{bai_bi-ventricular_2015, mauger_right_2019, burns_genetic_2024}, four-chamber cardiac anatomy in \citep{rodero_linking_2021}, and bi-atrial anatomy in \citep{nagel_bi-atrial_2021}). However, some limitations can be identified. The spatial resolution of the input data is often limited compared to the clinical state-of-the-art \citep{dodd_evolving_2023, bai_bi-ventricular_2015, mauger_right_2019, nagel_bi-atrial_2021, burns_genetic_2024}. Furthermore, some studies include only a limited and possibly insufficient number of subjects in their shape analysis, such as 19 shapes in \citep{rodero_linking_2021} and 47 shapes in \citep{nagel_bi-atrial_2021}. Such limitations are often related to the lack of automatic workflows for the generation of training geometries, resulting in time-consuming manual processes, especially in the segmentation step \citep{mauger_right_2019, rodero_linking_2021, nagel_bi-atrial_2021}. Moreover, some studies include subjects suffering from cardiovascular disease, meaning that the resulting SSM is likely not representative of a healthy population \citep{mauger_right_2019, nagel_bi-atrial_2021, burns_genetic_2024}. In terms of methodology, none of the studies leverage the recent advancements in the field of cardiac modeling, specifically designed for registration purposes, i.e. the implementation of universal cardiac coordinate systems \citep{bayer_universal_2018, roney_universal_2019,gillette_framework_2021, zappon_efficient_2025}. Additionally, although cardiac SSM studies usually include the release of synthetic cohorts, many of these cohorts only feature surface meshes \citep{bai_bi-ventricular_2015, mauger_right_2019}, and as such are not suitable for direct use in downstream applications, such as finite element simulations. Notable exceptions are \citep{rodero_linking_2021} and \citep{nagel_bi-atrial_2021}, which include the publication of volumetric four-chamber meshes and volumetric bi-atrial meshes, respectively. Lastly, while \citep{rodero_linking_2021} and \citep{nagel_bi-atrial_2021} have studied the relationship between anatomical features, defined by their cardiac SSM, and in silico functional measurements of the heart, the impact of anatomical differences on functional measurements in real-world patients, which have the power to either validate these simulation findings or indicate modeling flaws, remains understudied.

\subsection{Main contributions}
\label{subsec6}

In this work, we develop a biventricular SSM from high-resolution cardiac CT scans from 271 healthy individuals. Our work advances the state-of-the-art in cardiac statistical shape modeling and CDT technology by:

\begin{enumerate}

    \item \textbf{Dataset construction:} We use a large dataset comprising 271 high-resolution (submillimeter) CT images from a healthy, mixed-sex population, enhancing accuracy and anatomical detail of the SSM. Moreover, fully automated segmentation and anatomical model generation workflows are implemented to build the SSM training dataset. This automated anatomical twinning approach directly enables building patient-level, anatomical, biventricular CDTs.

	\item \textbf{Statistical shape modeling:} We establish lightweight correspondence between point clouds of distinct geometries, leveraging universal ventricular coordinates (UVCs) \citep{bayer_universal_2018}, effectively eliminating the computationally expensive nonlinear registration step typically required for SSM construction.

	\item \textbf{Providing high-quality meshes for downstream applications:} We provide a detailed biventricular SSM, together with a synthetic cohort of 100 ready-to-use biventricular meshes, including anatomical labels and fibers, representative of healthy biventricular anatomy. The virtual anatomical cohort represents a building block towards a population-level CDT.

	\item \textbf{Quantifying relationships with biventricular anatomy:} We study and quantify the relation between biventricular anatomy and clinically measured ECGs and demographic variables, leveraging statistical shape analysis and automated ECG delineation.
	
\end{enumerate}

\section{Methodology}
\label{sec2}

\subsection{Phase I: Dataset construction}
\label{phaseI}

\subsubsection{Raw input data}

This study combines data from the Master@Heart study \citep{de_bosscher_endurance_2021} with a proprietary dataset containing clinical data from female subjects, measured at the University Hospital of Leuven (UZ Leuven). All of these subjects were screened for absence of coronary artery disease. The Master@Heart study is a multicenter (University Hospitals Leuven, University Hospital Antwerp, and Jessa Hospital Hasselt) prospective cohort trial, comparing cardiovascular health in lifelong endurance athletes, athletes who started training later in life, and healthy non-athletic controls, (ClinicalTrials.gov Identifier NCT03711539). In the remainder of this text, these subgroups will be referred to as early-onset athletes, late-onset athletes, and healthy controls, respectively. The Master@Heart dataset covers male subjects between the ages of 45 and 70. The following cardiovascular risk factors acted as exclusion criteria for this study: a medical history of cardiovascular disease, current or past smoking, use of antidiabetic drugs, statins, or antihypertensive drugs or a body mass index (BMI) exceeding 27.2~$kg/m^2$. From this Master@Heart dataset, a subset, pertaining to subjects having their clinical measurements taken at UZ Leuven, was used. For consistency, female subjects of the same age group were selected from routine clinical care to complement this male-only dataset. Female subjects were excluded if they present with more than three of the following risk factors: smoking, arterial hypertension, diabetes, a family history of cardiovascular disease, and dyslipidemia. Given that this second dataset was obtained from clinical care rather than a dedicated clinical trial, the freedom in the selection process was more limited. However, the selection criteria are expected to select a population representative of healthy biventricular anatomy and function.

\begin{table}
\centering
\begin{tabular}{c c c}
\hline
\textbf{Subgroup} & \textbf{Sex} & \textbf{Number} \\
\hline
 Healthy controls & Male & 61 \\
 Early-onset athletes & Male & 66 \\
 Late-onset athletes & Male & 57 \\
 Unknown & Male & 4 \\
 Healthy & Female & 83 \\
 \hline
\end{tabular}
\caption{Number of subjects used in this study, grouped by lifestyle and sex.}\label{table_data}
\end{table}

For both male and female subjects, ECG-triggered CT coronary angiography scans were acquired using a 128-slice dual-source CT scanner (Siemens Somatom Force, Siemens Healthineers, Forchheim, Germany), with images captured in diastole. The image resolutions are within the ranges $[0.25\text{-}0.42] \times [0.25\text{-}0.42] \times 0.6$ \SI{}{\milli\meter} for male subjects and $[0.25\text{-}0.38] \times [0.25\text{-}0.38] \times 0.6$ \SI{}{\milli\meter} for female subjects. The exact number of CT scans per demographic group is provided in Table \ref{table_data}. Ten-second, 12-lead ECG recordings were recorded for all subjects, except for three male subjects. The electrical axis was automatically determined from these ECGs using the Marquette 12SL ECG Analysis Program (GE Healthcare). Demographic data, including weight, height, sex, and age, and lifestyle data, indicating the subgroup (male early-onset athletes, male late-onset athletes, male healthy controls, or female subjects) were collected for all subjects, except for four male subjects. Secondary metrics, namely body mass index and body surface area (BSA) were calculated from weight and height. BSA was derived using the Du Bois formula \citep{du_bois_formula_1989}.

\subsubsection{Semantic segmentation}
In this work, we employed a 3D Convolutional Neural Network (CNN) to obtain semantic segmentations of the heart from CT scans.
The CNN was implemented as a 3D U-Net-like architecture~\cite{Ronneberger2015-ih} and was trained using the CT data of the publicly available MM-WHS dataset~\cite{Zhuang2019-uj}.
During training, we synthetically diversified the dataset by employing the sophisticated data augmentation toolkit by~\cite{Payer2018,payer2019integrating}, as well as random convolution-based data augmentation similar to~\cite{Xu2020-vo,Ouyang2023-xf,Choi2023-qy}.
Due to GPU memory constraints, all data was resampled to a physical resolution of $1.5 \times 1.5 \times 1.5$ \SI{}{\milli\meter} before being processed by the CNN.
Afterward, each prediction was resampled to its original resolution.
The resulting segmentation provides distinct anatomical labels for the left ventricular (LV) blood pool, LV wall, right ventricular (RV), right atrial, left atrial blood pool, aorta, and pulmonary arteries.
All automatically generated segmentations were assessed by experts and manually corrected whenever necessary.

\subsubsection{Anatomical model generation and UVC computation}
\label{uvcs}
For the generation of volumetric biventricular anatomical models \citep{crozier2016image}
and UVCs \citep{bayer_universal_2018} 
we employed the bi-ventricular model generation tool from \href{https://numericor.at/rlb/wordpress/products/}{NumeriCor CardioTwin} (NumeriCor GmbH, Graz, Austria).
Briefly, starting with automatically segmented anatomical labels, a series of rules and image processing operations were implemented to grow the RV wall and re-label the LV wall.
RV walls are thinner than their LV counterparts, and accurately segmenting them using current routine clinical imaging modalities, such as CT and MRI, is often not feasible \cite{kawel2020reference}. To this end, our workflow incorporated prior knowledge \cite{bhattacharya2024right, kawel2020reference} to define the RV wall thickness, as well as the distribution of endocardial and epicardial tissue in both ventricular walls. Image operations were applied to the image stack, including inward and outward extrusion, to generate the RV wall from the blood pool surface. For the LV, starting with the already segmented endocardial surface, an extrusion process was employed to assign new labels for the endocardial and epicardial tissues. Extrusion of the walls was restricted at the interfaces between the LV and left atrial blood pool, the LV blood pool and aorta, the RV and right atrial blood pool, and the RV blood pool and pulmonary arteries. These restrictions effectively created orifices corresponding to the mitral valve (MV), aortic valve (AV), tricuspid valve (TV), and pulmonary valve (PV). During this phase, the valve rings were additionally labeled with a specified thickness.

Following the extrusion phase, a volumetric anatomical mesh was generated by first extracting a labeled surface mesh encompassing all ventricular wall and valve labels. This initial surface mesh, conforming to the jagged voxel representation, underwent several remeshing and smoothing steps to achieve a smooth representation of all ventricular walls. This process included ensuring smooth transitions between wall segments of varying widths, applying topological corrections, and performing quality checks and enhancements to ensure topological consistency and high mesh quality. Surface meshes were resampled to a resolution of 0.9 \SI{}{\milli\meter}, making them suitable for electrophysiological simulations using reaction-eikonal models \citep{neic_efficient_2017,gillette_framework_2021}. In the final stage, volumetric meshes of the biventricular walls were created by inward meshing of the labeled surface meshes. This step was followed by additional topological corrections, reindexing, refinement, and mesh quality enhancement procedures. The resulting biventricular anatomy is represented as an unstructured volumetric tetrahedral mesh with fundamental anatomical labels, including endocardial and epicardial domains, as well as the MV, TV, AV, and PV regions.

\begin{figure*}[!t]
	\centering
	\includegraphics[width=0.75\textwidth]{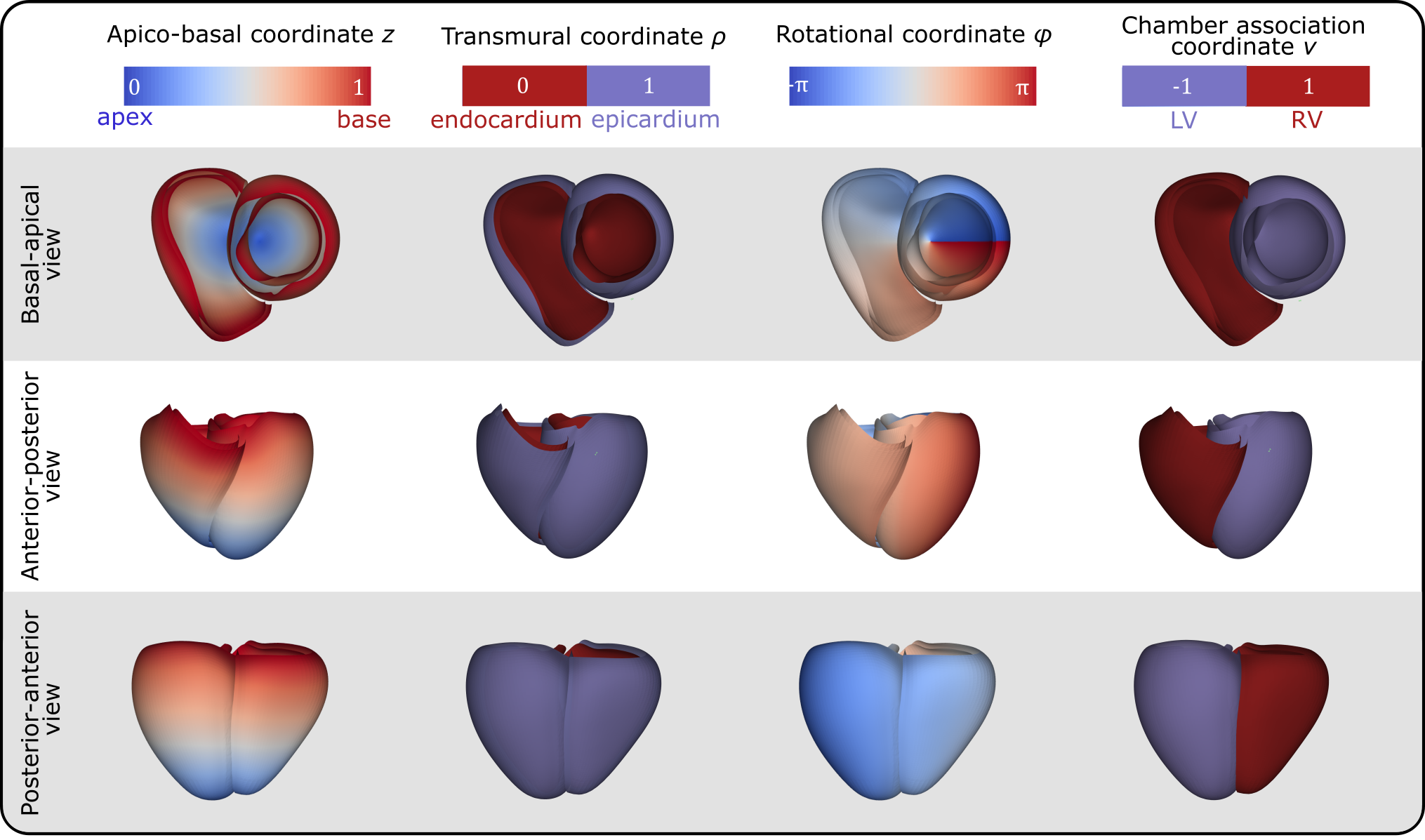}
	\caption{Overview of the UVCs, with important ventricular structures and viewpoints used throughout this paper.}
	\label{fig:general}
\end{figure*}

Four universal ventricular coordinates (UVCs) were automatically computed, as described in \citep{gillette_framework_2021}, to anatomically characterize positions within the ventricles. The computed UVCs include: (1) a long-axis coordinate $z \in [0,1]$, spanning from the apex to the ventricular base, thus also called apico-basal; (2) a transmural coordinate $\rho \in [0,1]$, extending from the endocardium to the epicardium; (3) a rotational coordinate $\varphi$ with $\varphi_{LV} \in [-\pi, \pi]$ for the LV and $\varphi_{RV} \in [\sfrac{-\pi}{2}, \sfrac{\pi}{2}]$ for the RV, capturing regions such as septal, infero-septal, inferior, lateral, anterior, and antero-septal; and (4) a chamber association coordinate $\nu \in \{-1,1\}$, indicating the LV or RV (we refer to Figure~\ref{fig:general} for a graphical representation of the UVCs).

The entire workflow was implemented using the \texttt{meshtool} package \citep{neic2020automating}, in combination with \texttt{Python} scripts. For meshing operations, the licensed software TetGen (Weierstrass Institute, Berlin, Germany) and NetGen (CerbSim GmbH, Vienna, Austria) have been integrated with \texttt{meshtool}. 
Interactive control and, if required, correction steps are fully supported 
by the freely available basic version of \href{https://numericor.at/rlb/wordpress/resources/}{CARPentry Studio Core} (NumeriCor GmbH, Graz, Austria).

\subsection{Phase II: Statistical shape modeling}

\subsubsection{Correspondence establishment and surface construction}
\label{surface_construction}

\begin{figure}[!t]
	\centering
	\includegraphics[width=0.5\textwidth]{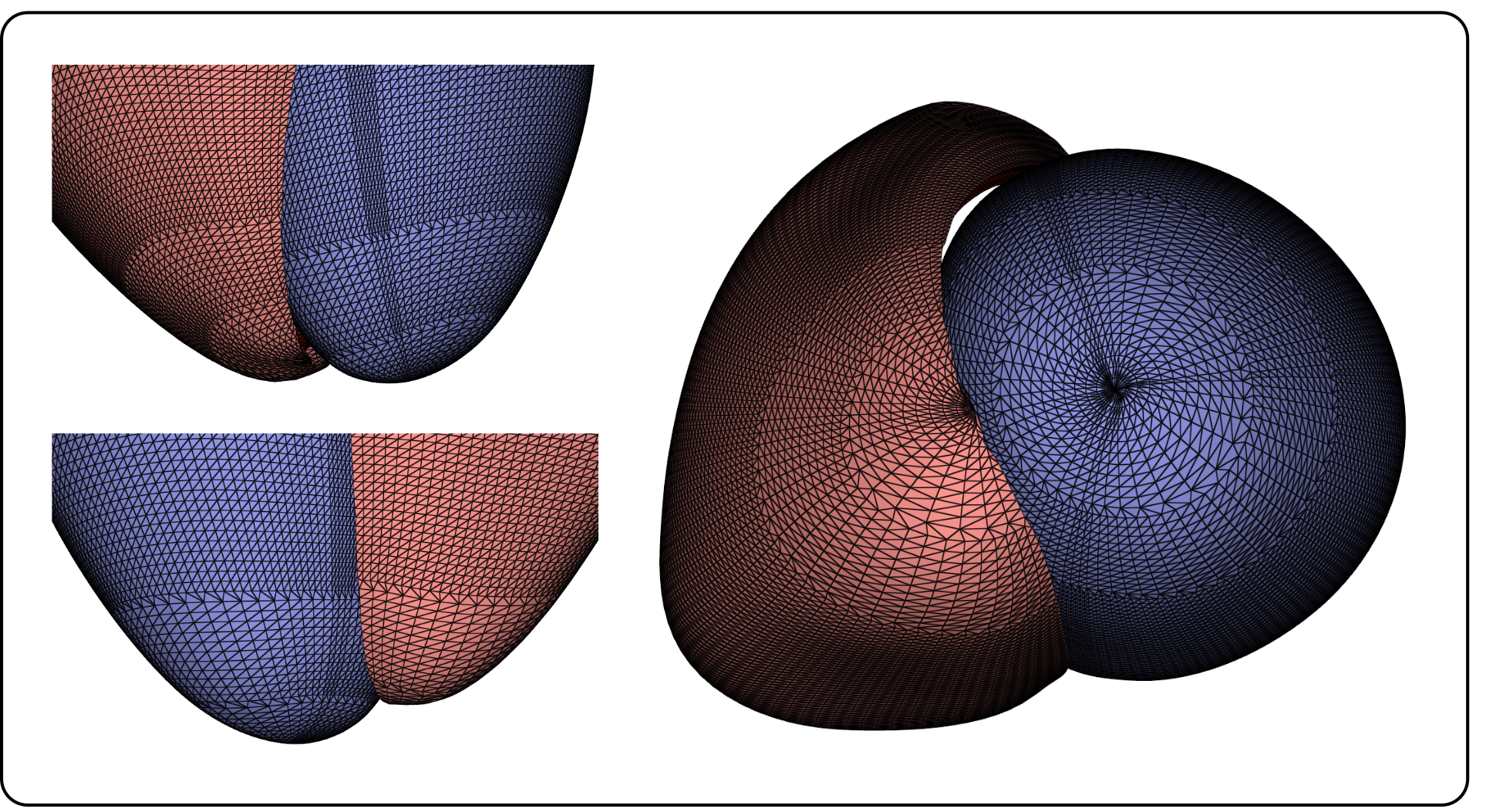}
	\caption{Apical coarsening shown in various views on the ventricles. Top left: anterior-posterior view, bottom left: posterior-anterior view, right: apical-basal view.}
	\label{fig:apical_coarsening}
\end{figure}

\begin{figure}[!t]
	\centering
	\includegraphics[width=0.5\textwidth]{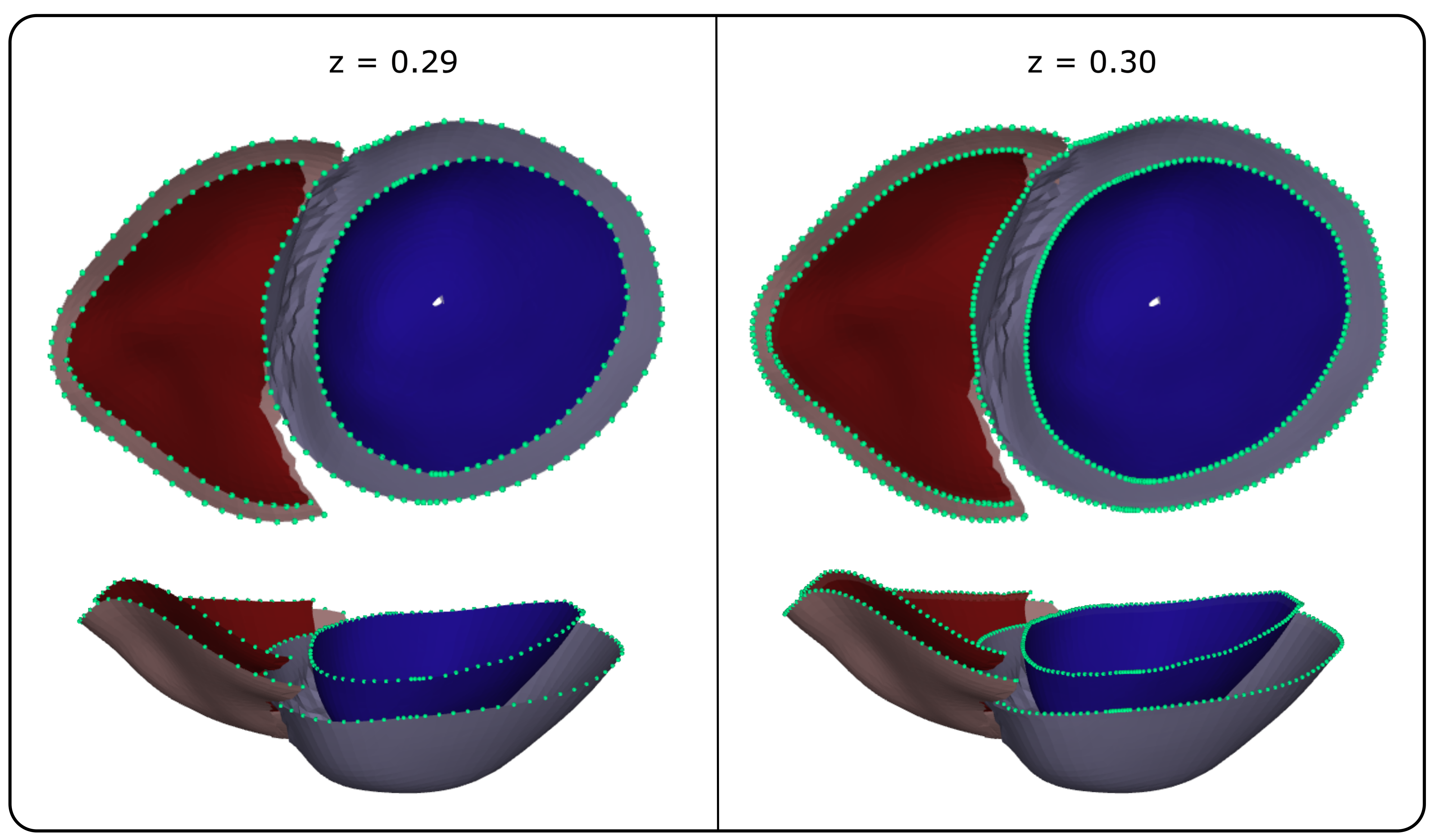}
	\caption{Graphical representation of the UVC-based point extraction process. The points extracted at two subsequent $z$ values are shown, using a different number of $\varphi$ values due to apical coarsening. Top: basal-apical view, bottom: anterior-posterior view.}
	\label{fig:point_extraction}
\end{figure}

For PCA, each anatomical geometry must be represented as a single feature vector, and comparability across different vectors, i.e. geometries, was established by means of corresponding point extractions from the original meshes. We tackled corresponding point extraction employing a new, lightweight method based on UVCs correspondence. To this end, a uniform sampling over the apico-basal $z$, and rotational $\varphi$ coordinate was applied to define a reference set of tuples ${(z_i, \rho \in \{0, 1\}, \varphi_j, \nu \in \{-1, 1\})}$. For both LV and RV, these tuples define a grid of corresponding points to be extracted from the endocardial ($\rho = 0$) and epicardial ($\rho = 1$) surfaces of the biventricular geometries. We sampled 80 values between $[0,1]$ for the apico-basal coordinate $z$. For the rotational coordinate $\varphi$, we selected 160 values from the interval $\left]-\pi,\pi\right[$ for the LV, and 80 values in the interval $\left]\sfrac{-\pi}{2},\sfrac{\pi}{2}\right[$ for the RV. A coarsening scheme was applied near the LV and RV apex, reducing the sampling density of $\varphi$ to avoid having an increasingly dense point cloud in Cartesian space due to smaller ventricular circumference in the apical region, as illustrated in Figure~\ref{fig:apical_coarsening}. When $z<0.30$, the sampling density of $\varphi$ was decreased by factor 2, resulting in the selection of 80 values for the LV, and 40 for the RV. For $z<0.15$, this sampling density was decreased by factor 4, selecting 40 and 20 values for the LV and RV, respectively. The sampling scheme resulted in point clouds consisting of $M = 31~724$ points, comparable to the number of points included in meshes of similar studies \citep{bai_bi-ventricular_2015, nagel_bi-atrial_2021}. The point extraction process is illustrated in Figure~\ref{fig:point_extraction}. For the rationale behind selecting this sampling scheme, we refer to \ref{app_sampling}.

For each tuple, a corresponding point was extracted from each geometry using the \texttt{meshtool} package \citep{neic2020automating}. Such a point can already exist as a point of the anatomical mesh, or as a point inside one of the elements of the mesh. In the former case, the point was extracted, while in the latter case, the point was computed by interpolation of the universal coordinates inside the corresponding mesh element. This approach was used to extract a point cloud in Cartesian coordinates for each anatomical geometry.

\begin{figure}[!t]
	\centering
	\includegraphics[width=0.4\textwidth]{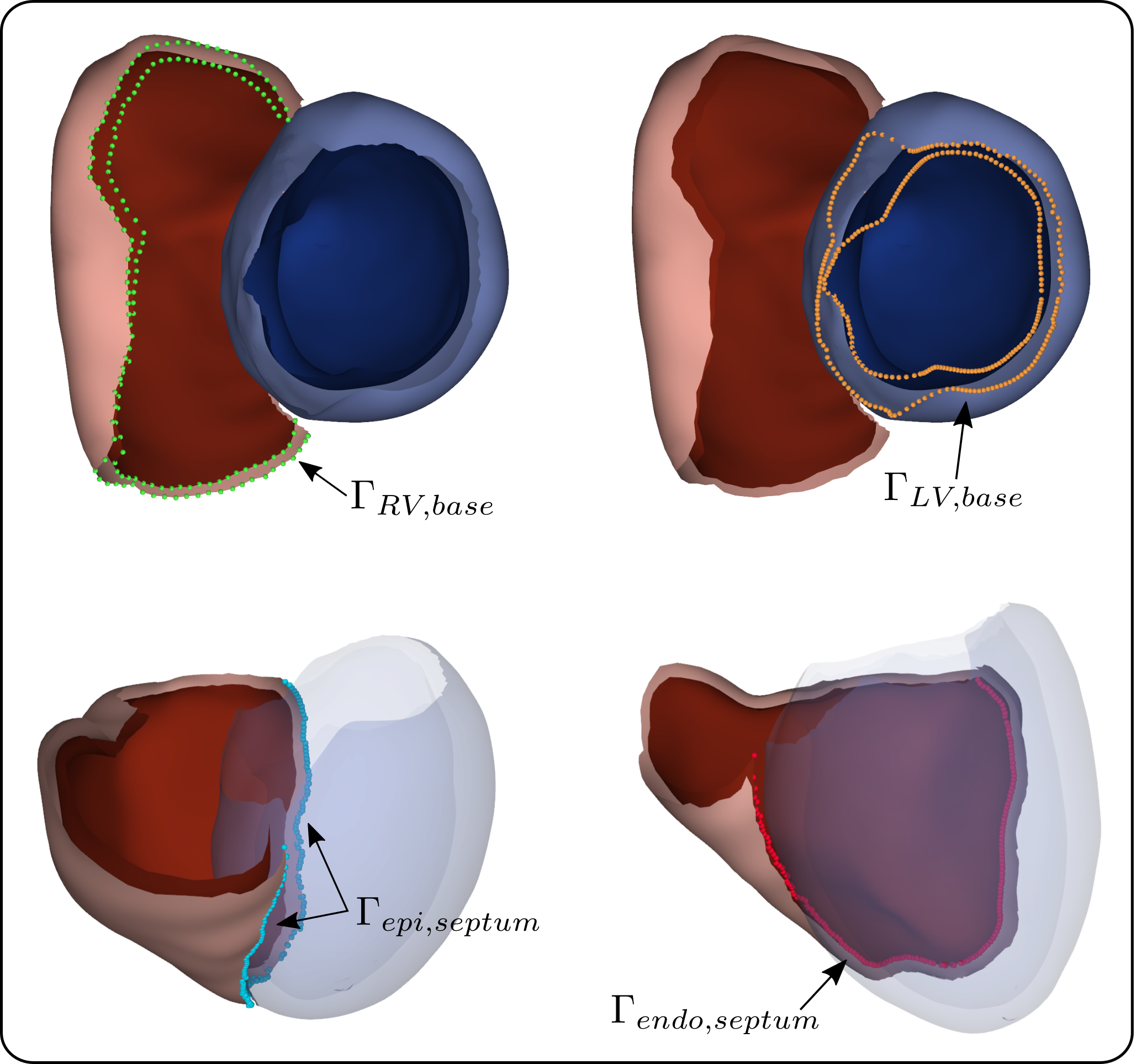}
	\caption{Graphical representation of the set of points extracted at the base of the RV $\Gamma_{RV,base}$, and LV $\Gamma_{LV,base}$, and of the RV points on the endocardium $\Gamma_{endo,septum}$ and epicardium $\Gamma_{epi,septum}$, to be connected to the LV.}
	\label{fig:closing_interfaces}
\end{figure}

In this stage, we moreover identified and extracted the set of points 
$(z_i = 1, \rho_i \in \{0, 1\}, \varphi_i, \nu = -1)$ as $\Gamma_{LV,base}$ and the points $(z_i = 1, \rho_i \in \{0, 1\}, \varphi_i, \nu = 1)$ as $\Gamma_{RV,base}$, effectively representing the ventricular bases. 
Additionally, for all discrete apico-basal levels $z_h$ we extracted the points $(z_i = z_h, \rho \in \{0, 1\}, \varphi_i, \nu = 1)$ of the RV for which $\varphi_i$ is maximal or minimal. For $\rho = 0$ the set is referred to by $\Gamma_{endo,septum}$ and for $\rho=1$ by $\Gamma_{epi,septum}$.
$\Gamma_{endo,septum}$ and $\Gamma_{epi,septum}$ then correspond to the sets of points on the epi- and endocardium of the RV to be connected to the epicardium of the LV (see Figure~\ref{fig:closing_interfaces} for a graphical representation).
Identification of these points was necessary to subsequently close the biventricular surface and generate a volumetric mesh.

\begin{figure}[!t]
	\centering
	\includegraphics[width=0.4\textwidth]{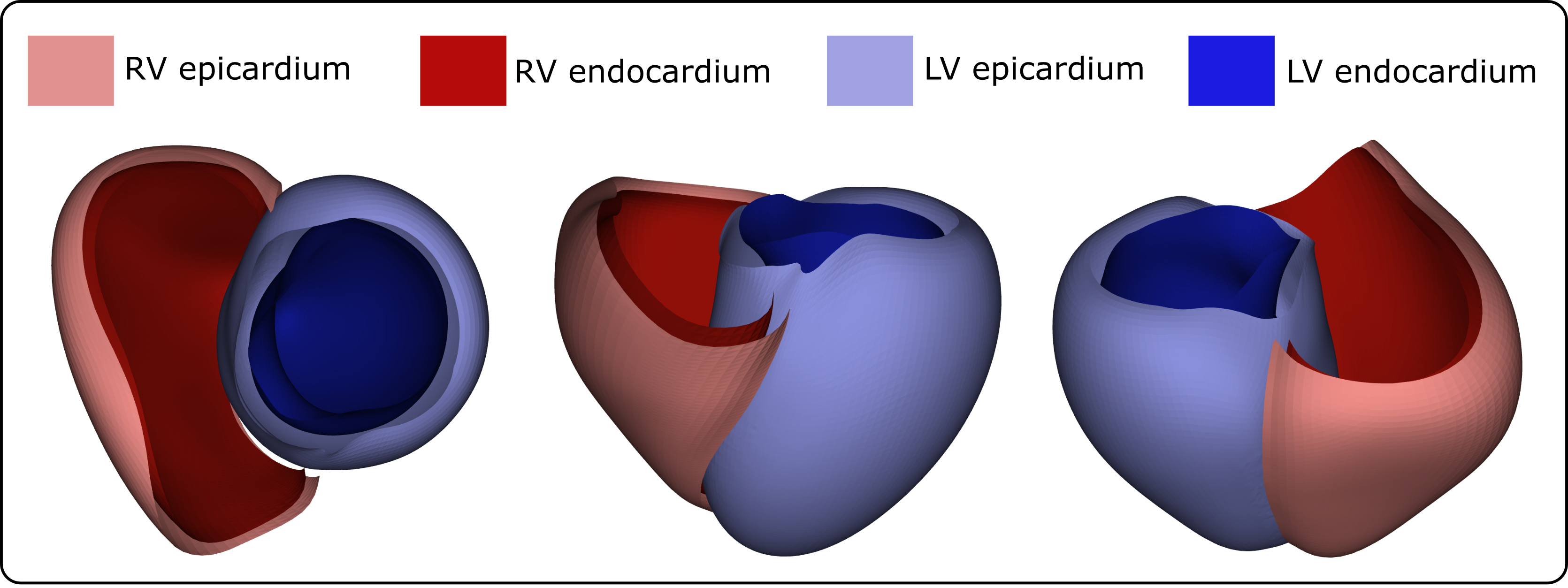}
	\caption{Generated endo- and epicardial LV and RV surfaces, shown in basal-apical, anterior-posterior and posterior-anterior view.}
	\label{fig:surfaces}
\end{figure}

The structured UVC-based sampling allowed for the straightforward generation of a 2D triangulation. Moreover, a minimal set of labels was automatically included in the triangulation, consisting of four distinct surfaces: LV endo- and epicardium and RV endo- and epicardium, as shown in Figure~\ref{fig:surfaces}. Per UVC construction, the septal surface of the RV was labeled as epicardium of the LV.

\begin{figure}[!t]
	\centering
	\includegraphics[width=0.48\textwidth]{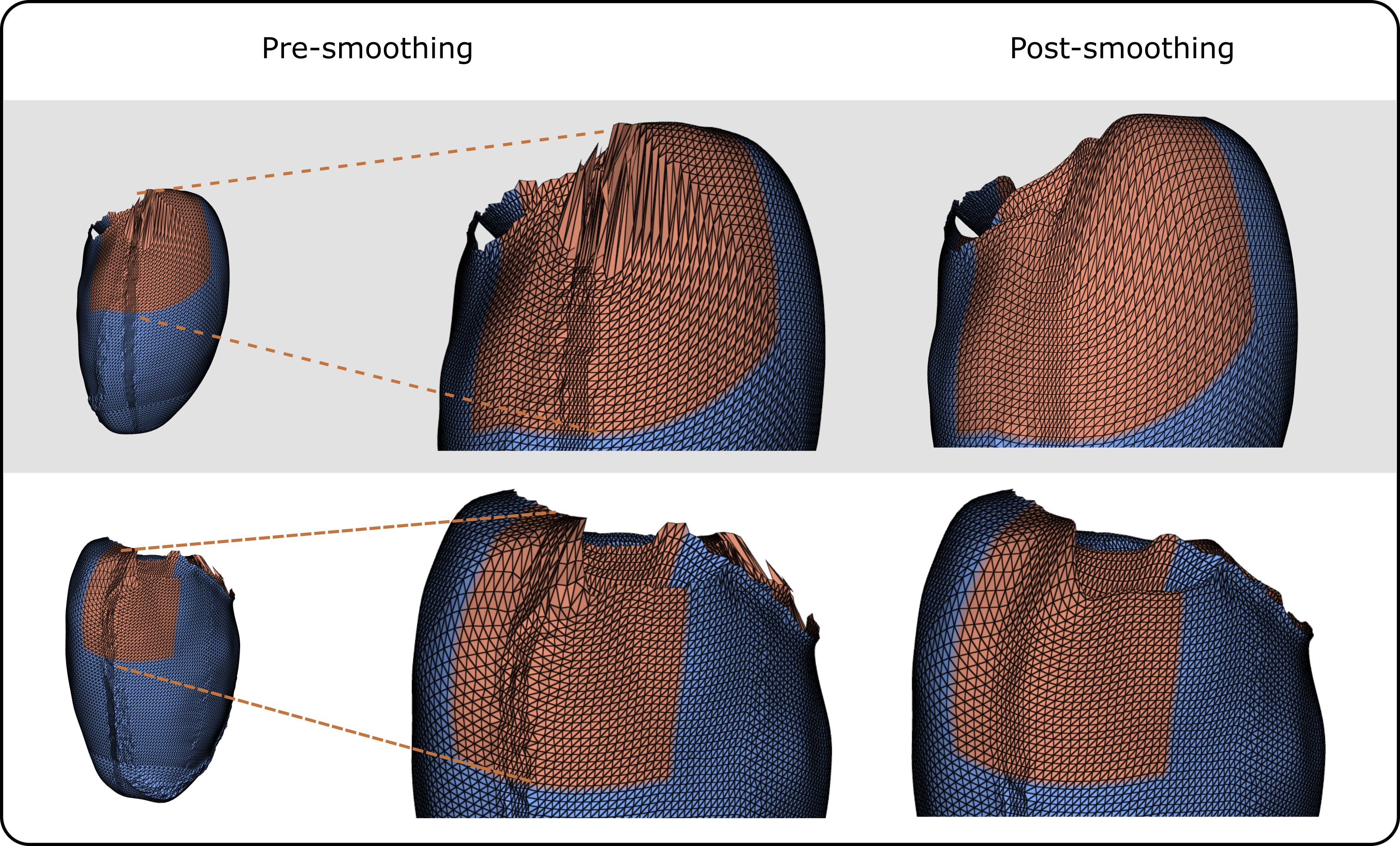}
	\caption{Effect of smoothing on the LV. The regions where smoothing is applied, are shown in orange. The top row shows the smoothing in the region where $\sfrac{\pi}{4} \leq \varphi_{LV} \leq \sfrac{3\pi}{4}$, the bottom row in the region where $\sfrac{-9\pi}{16} \leq \varphi_{LV} \leq \sfrac{-3\pi}{16}$.}
	\label{fig:smoothing}
\end{figure}

To mitigate artifacts within the resampled surface meshes arising from the irregular distribution of UVCs in the Cartesian space, Laplacian smoothing \citep{preim_chapter_2014} was applied at two specific mesh regions, selected based on visual inspection of several anatomical meshes. To maintain pointwise correspondence, the same smoothing scheme was uniformly applied to all surface meshes. Smoothing was applied on the LV endo- and epicardium, in two regions. The first region is defined by $0.70 \leq z \leq 1$, and $\sfrac{\pi}{4} \leq \varphi_{LV} \leq \sfrac{3\pi}{4}$. The second region is defined by $0.70 \leq z \leq 1$ and $\sfrac{-9\pi}{16} \leq \varphi_{LV} \leq \sfrac{-3\pi}{16}$. The results of the smoothing operations are illustrated in Figure~\ref{fig:smoothing}. 

\subsubsection{Generalized Procrustes alignment of point clouds}

The point clouds from all subjects were aligned using a generalized Procrustes method \citep{gower_generalized_1975}, an iterative technique that progressively moves the set of geometries, minimizing the Frobenius norm across all point clouds. In this iterative procedure, point clouds were translationally and rotationally aligned, while scaling was impeded, as size factors are intended to be captured in the SSM. In a parallel analysis, we performed a Procrustes alignment where scaling was allowed, to study heart shape effects independently of heart size, as explained in \ref{app0}. The initial reference point cloud for the iterative procedure was randomly chosen from the set of point clouds, while it was defined as the average of all point clouds in each subsequent iteration. Rotation matrices for pairwise alignment of point clouds to the reference were calculated with the algorithm proposed by Schonemann \citep{schonemann_generalized_1966}, while 
translational movements were obtained from point cloud centering. The iterative procedure was automatically stopped at the stabilization of the average of all aligned point clouds, corresponding to a computed Frobenius norm between subsequent averages smaller than a prescribed tolerance $\xi$, which was set in this work as $10^{-5}$.

\subsubsection{SSM construction and evaluation}
\label{eval}

Each aligned point cloud $Y^i \in\mathbb{R}^{M\times3}$ was first vectorized to the form $\tilde{Y}^{i}\in\mathbb{R}^{1\times3M}$. Row-wise concatenation of all vectorized, aligned point clouds resulted in a feature matrix $F \in \mathbb{R}^{N \times 3M}$. The feature matrix was centered by subtraction of the computed mean geometry, and PCA was applied, leading to a set of $(N-1)$ principal components (PCs) $C \in \mathbb{R}^{(N-1)\times 3M}$, and a monotonically decreasing set of variances associated with each of these PCs $\left[\lambda_1, \ldots , \lambda_{N-1}\right]$. 

The PCs can be viewed as a new spatial reference system, an alternative to the vectorized point cloud representation of the biventricular geometries. Consequently, a biventricular mesh can be represented in the PC system as a weighted linear combination of the PCs, where the set of weights, called PC scores, provides a dimensionally reduced representation of the geometry in the PC space. 

A biventricular geometry, described as a vectorized point cloud in Cartesian coordinates as per the training set, can thus be transformed into its low-dimensional representation through the matrix $C^T$, and conversely, a low-dimensional representation can be transformed back into a point cloud using $C$. This process is particularly true for the geometries $Y^i$ used in the PCA training set, which can be fully reconstructed in both directions. Furthermore, if the matrix $C$ retains all PCs, the reconstruction is exact. When some PCs are excluded, the reconstruction becomes an approximation, and the associated reconstruction error can be quantified. Typically, such approximations are obtained by excluding the final $L$ PCs, with $L < N-1$, which are associated with lowest eigenvalues, and therefore encode the smallest amount of information.

Moreover, the PCs can be used to approximate biventricular geometries outside the training set. For a new biventricular geometry, after applying the UVC-based sampling method used to generate the training dataset, it is possible to apply $C$ to approximate the geometry and compute the corresponding reconstruction error. This error serves as a measure of the SSM capability to generalize to previously unseen geometries.

The reconstruction error was defined as the mean distance between corresponding points in the original and reconstructed point clouds, identically to \citep{nagel_bi-atrial_2021}. Averaging multiple reconstruction errors across subjects provides the overall reconstruction error associated with an SSM.

To investigate the reconstruction error of the SSM across different groups -- namely, the female group, the male control group, male early-onset athletes, and male late-onset athletes -- a leave-one-out cross-validation (LOOCV) experiment was performed over the complete dataset on the one hand, and a stratified dataset, containing 56 subjects from each of the four demographic subgroups, on the other hand. The two experiments consisted of 271 and 224 iterations, respectively, equal to the total number of geometries in the dataset of the experiment. In each iteration, a single geometry was removed from the training set, and a new SSM was built from the remaining geometries. This SSM was consequently used to reconstruct the left-out geometry, and the resulting reconstruction error was calculated. Comparing the reconstruction errors across the various demographic groups provides a notion of the suitability of the SSM to represent certain types of anatomies within this study population.

To investigate the number of geometries needed for effective SSM construction, we analyzed the reconstruction error across SSMs with varying training set sizes. Four training sets were constructed, containing either 56, 112, 168, or 224 geometries, randomly sampling an equal amount of geometries from each demographic group without replacement. A fifth dataset contained all 271 available geometries. A separate LOOCV experiment was carried out within each of these five datasets.

Furthermore, we tested the hypothesis that male and female cardiac anatomy differ significantly \citep{st_pierre_sex_2022}, by investigating the bias introduced by a sex-specific (biased) SSM. We quantified this bias as the reduced ability to represent geometries of the opposite sex compared to a mixed-sex (unbiased) SSM. To this end, we built two biased SSMs, one trained on 60 male subjects from the control group, and one trained on 60 healthy female subjects, in both cases randomly selected. We also trained an unbiased SSM on a random selection of 30 female and 30 male subjects, and we compared the main modes of variability of the three SSMs. For each dataset, we performed a LOOCV experiment to establish the baseline representation ability of each SSM. Moreover, we calculated the reconstruction error on geometries not present in the training set for each SSM. These were mixed-sex geometries for the unbiased SSM, and subjects from the opposite sex for both biased SSMs.

\subsection{Phase III: Synthetic cohort generation}

\subsubsection{Synthetic surface mesh generation}

For the construction of synthetic meshes, we sampled values for the first 94 PCs, which encoded 99\% of the total variance in the population. Under the assumption that the PC scores are normally distributed for geometries in the training set, synthetic geometries can be sampled from the SSM by drawing samples from the following distribution for each PC $k \in {1,...,94}$:

\begin{equation}
s_k \sim \mathcal{N}(0,\lambda_k)
\end{equation}

We verified the underlying assumption of normality in \ref{app3} for the first 16 PCs.

The resulting PC scores were combined into a set of vectors $Y^i_{PC} \in \mathbb{R}^{1 \times (N-1)}$, corresponding to the low-dimensional representations of new geometries. Each vector was then transformed into a vector of dimension $3M$ using matrix $C$, and subsequently reconstructed as a regular point cloud in Cartesian coordinates, representing the synthetic geometry. The points in the new point clouds retain the same ordering as those in the point clouds used to train the SSMs. Consequently, the triangulation applied during the training set sampling was reused on the synthetic point clouds, effectively generating new synthetic meshes.

In rare cases, the synthetic meshes displayed non-physiological characteristics. Two categories of such characteristics were identified: cases where the endo- and epicardium intersected in the LV apical region, and cases where they intersected in the septal region. These synthetic meshes were replaced by sampling new instances from the SSM. Apart from actual intersections, also cases where the minimal distance between the endo- and epicardium in the apical or septal region was too small to be accurately represented with the chosen mesh resolution, were replaced. In total, eight synthetic meshes were replaced in this process.

\subsubsection{Volumetric mesh and fiber generation}
\label{surface_closing}
The synthetic meshes are surface bilayer biventricular models opened at the RV and LV junctions and at the biventricular bases. Consequently, the SSM synthetic models are not directly suitable for volumetric meshing. To address this, two distinct closing operations were implemented: one for the RV and LV junction and another for the ventricular bases.

The closure of the RV to LV junction was achieved by connecting the points in $\Gamma_{epi,septum}$ and $\Gamma_{endo,septum}$ to the nearest point on the LV epicardium, thereby defining the set of points $\Gamma_{septum}$. A triangulation was then computed between the RV and LV points. To prevent the creation of holes in the resulting meshes, geodesic paths were generated to connect any two consecutive points in $\Gamma_{septum}$ that were not adjacent. The points along these paths were extracted and added to $\Gamma_{septum}$. During this process, the RV septum was also tagged based on the UVC coordinates of the points in $\Gamma_{septum}$.

For the closure of the ventricular bases, the uniform UVC sampling across the endocardium and epicardium of each chamber ensured that the same number and order of points could be identified on both surfaces. Consequently, corresponding points on the endocardium and epicardium of each chamber were simply connected to build the triangular surfaces of the RV and LV bases.

After the closing phase, a volumetric mesh was generated using the flood fill algorithm of the \texttt{fTetWild} library \citep{Yixin2020}. This procedure not only generates a tetrahedral volumetric mesh within the provided closed surface but also performs mesh cleaning by resolving potential mesh intersections. However, anatomical labels were not preserved during the meshing phase, and the resulting volumetric mesh size was comparable to the coarse grid obtained with the UVC sampling, therefore still not suitable for EP simulations.

The generated volumetric mesh was subsequently resampled to a resolution of \SI{0.9}{\milli\meter}, and the anatomical labels were projected onto the endocardial and epicardial surfaces of the refined volumetric mesh, then interpolated throughout the volume. Finally, using the method described in \cite{gillette_framework_2021}, a fiber architecture was generated following the rule-based approach proposed by Bayer et al. in \cite{bayer2012novel}. The entire workflow -- including mesh resampling, label projection and interpolation, boundary condition extraction, and fiber generation -- was fully automated using the \texttt{meshtool} software \cite{neic_efficient_2017}.

\subsection{Phase IV: Quantification of relationships with biventricular anatomy}

\subsubsection{Extraction of ECG features}
\label{ecgfeatures}

Quantitative features were extracted from clinical 10-second, 12-lead ECG signals to facilitate downstream statistical analyses of the SSM. We used a fully automated approach for ECG preprocessing and feature extraction, leveraging the \texttt{MATLAB} toolbox \texttt{ECGdeli} \citep{pilia_ecgdeli_2021}. Two types of features were collected: interval-related features, which pertain to the timings of the various waves in the ECG signals, and amplitude-related features, which correspond to the peak heights of the different ECG waves. The selection of these features was guided by their well-established relevance in the clinical assessment of cardiovascular health \citep{thygesen_universal_2007, chetran_ecg_2022}.

All ECG signals were pre-processed using a series of filtering steps from \texttt{ECGdeli}: a high-pass Butterworth filter with a cutoff frequency of 0.3 Hz to remove baseline wander, a low-pass Butterworth filter with a cutoff frequency of 120 Hz to eliminate high-frequency noise, and an isoline correction step, necessary for subsequent extraction of amplitude features. A complete list of the considered peak intervals and wave amplitudes, together with the corresponding leads, are listed in Table \ref{table:features}. For all interval-related features, the results from all 12 leads were consolidated, whereas amplitude-related features were calculated separately for single leads. For calculating the corrected QT interval, the Fridericia correction was implemented \citep{fridericia_systolendauer_1920}.

\begin{table}
\centering
\begin{tabular}{c c c}
\hline
\textbf{Feature name} & \textbf{Feature meaning} & \textbf{Source lead(s)}\\
\hline
 Pdur & P wave duration & 12 leads \\
 QRSdur & QRS duration & 12 leads \\
 Tdur & T wave duration & 12 leads \\
PQint & PQ interval & 12 leads \\
PRint & PR interval & 12 leads \\
QTint & QT interval & 12 leads \\
QTcint & Corrected QT interval & 12 leads \\
RRint & RR interval & 12 leads \\
\hline
Pamp II & P wave amplitude & lead II \\
Ramp II & R wave amplitude & lead II \\
Samp II & S wave amplitude & lead II \\
Tamp II & T wave amplitude & lead II \\
Ramp V6 & R wave amplitude & lead V6 \\
Samp V6 & S wave amplitude & lead V6 \\
Tamp V6 & T wave amplitude & lead V6 \\
 \hline
\end{tabular}
\caption{Overview of the ECG-derived features. The features are grouped in interval-related and amplitude-related features. While wave intervals were extracted from all leads, amplitude-related features were computed from specific leads, also reported in the table.}\label{table:features}
\end{table}

\subsubsection{Statistical testing}

As a downstream analysis of the SSM, we conducted statistical tests to explore relationships between subject-specific PC scores and additional variables. Our first analysis targeted the relations between biventricular anatomy and ECG through a statistical correlation analysis, yielding quantitative insights into the interplay between cardiac anatomy and EP. In a second analysis, we examined the association between biventricular anatomy and demographic factors.

To quantify the relationship between biventricular anatomy and cardiac EP, we investigated statistical correlations between PC scores and ECG-derived features. The 16 ECG-derived features used in this analysis included those detailed in Section \ref{ecgfeatures}, and the electrical axis of the heart. Pearson correlation coefficients were calculated to assess the strength and direction of the observed correlations \citep{pearson_1895}. Removal of outliers was unnecessary, as no outliers -- defined as observations that were more than three standard deviations away from the mean for either the PC scores or the ECG features -- were present. To verify statistical significance, two-tailed t-tests were performed. For the identification of correlations between anatomy and numerical demographic variables, namely height, weight, age, BMI, and BSA, the same statistical analyses was performed, again without any observed outliers.

To explore the relationship between biventricular anatomy and categorical demographic variables, mean differences between PC scores across demographic groups were computed. More specifically, the following comparisons were made: female subjects vs. male healthy controls, male healthy controls vs. male early-onset athletes, male healthy controls vs. male late-onset athletes, and male early-onset athletes vs. male late-onset athletes. The statistical significance of mean differences was evaluated using two-tailed t-tests.

The described analyses lead to a total of 25 tests per PC. To limit the total number of statistical tests, we focused on the first three PCs, as these captured the largest portions of the anatomical variance within the dataset. Consequently, a total number of $3\cdot25=75$ statistical tests were performed. To control the family-wise error rate at a threshold of 0.05, we apply the Holm-Bonferroni correction for multiple testing \citep{holm_simple_1979}. 

\section{Results}
\label{sec3}

\subsection{Evaluation of the full SSM}

\begin{figure}[!t]
	\centering
	\begin{subfigure}{0.48\textwidth}
		\centering
		\includegraphics[width=\textwidth]{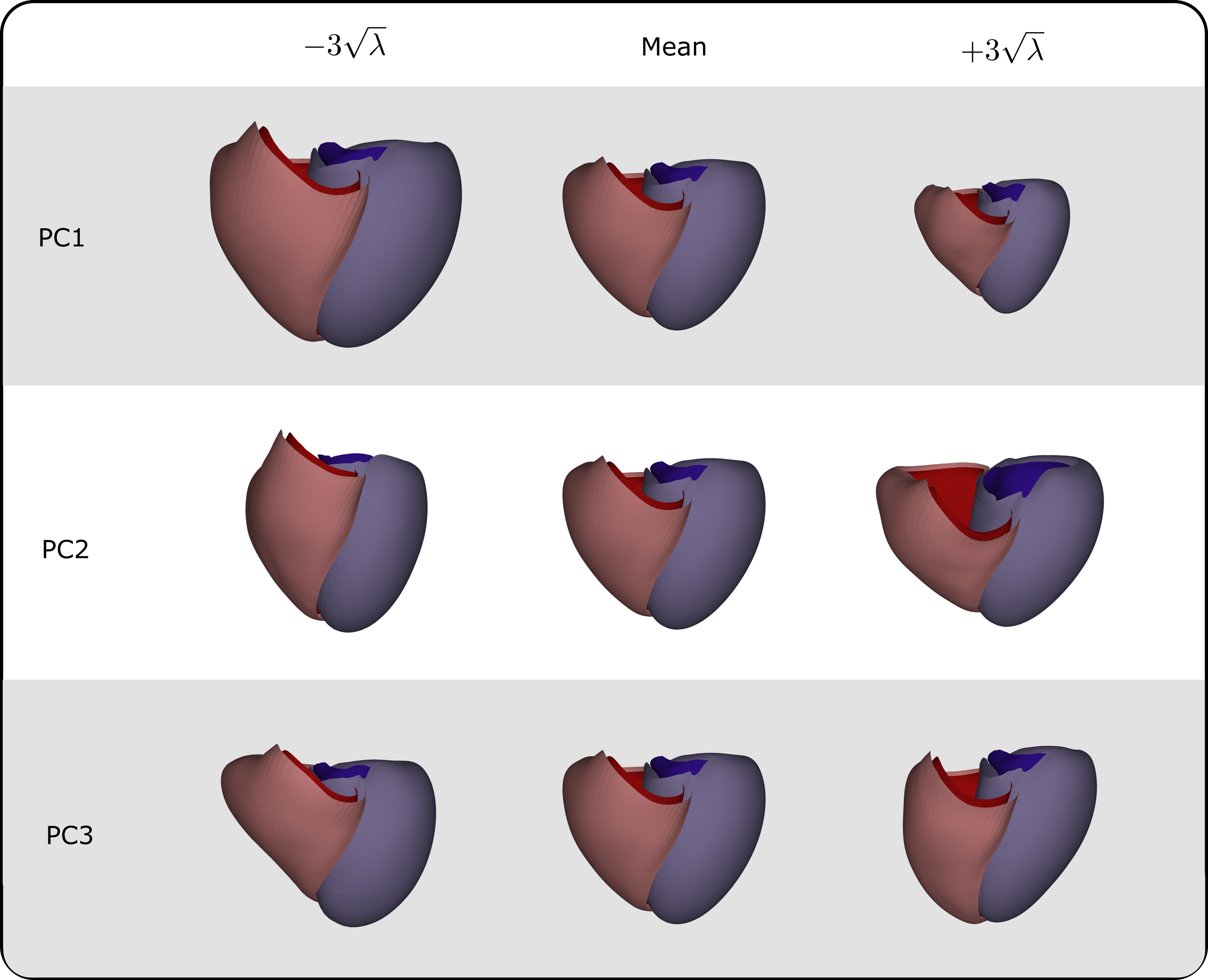}
		\caption{\footnotesize{Anterior-posterior view of the first three PCs.}}
		\label{fig:fullSSM_front}
	\end{subfigure}
	\hfill
	\begin{subfigure}{0.48\textwidth}
		\centering
		\includegraphics[width=\textwidth]{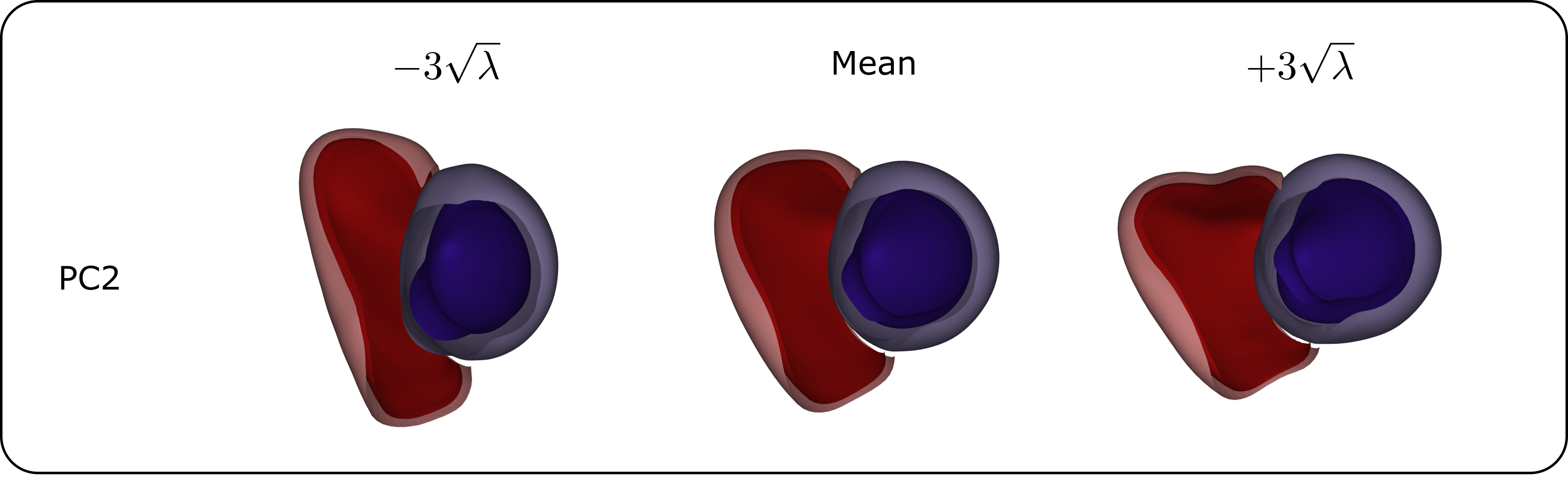}
		\caption{\footnotesize{Basal-apical view on PC2.}}
		\label{fig:SSM_PC2}
	\end{subfigure}
	\caption{Overview of the statistical shape model, trained on the unbiased study population. The effect of each principal component $\text{PC}_i$ is shown by setting its score to $\text{-}3\sqrt{\lambda_i}$, 0 or $\text{+}3\sqrt{\lambda_i}$.}
	\label{fig:fullSSM}
\end{figure}

The SSM, trained on the complete dataset of 271 biventricular geometries, achieves substantial dimensionality reduction. Specifically, the first 3 PCs account for 40.34\%, 11.19\%, and 9.59\% of the shape variability in the population, respectively. The first 16, 29, and 94 PCs collectively explain 90\%, 95\%, and 99\% of the biventricular shape variability, respectively. The primary modes of variation encoded by the first few PCs are as follows: the first PC represents overall ventricular size, the second PC captures the shape and elongation of the ventricles, with the RV showing the most notable changes, and the third PC describes the relative positioning between the LV and RV, along the apical-basal axis (see Figure~\ref{fig:fullSSM} for detailed views). Additional PCs are mostly connected to RV shape.

\begin{figure}
	\centering
	\includegraphics[width=0.48\textwidth]{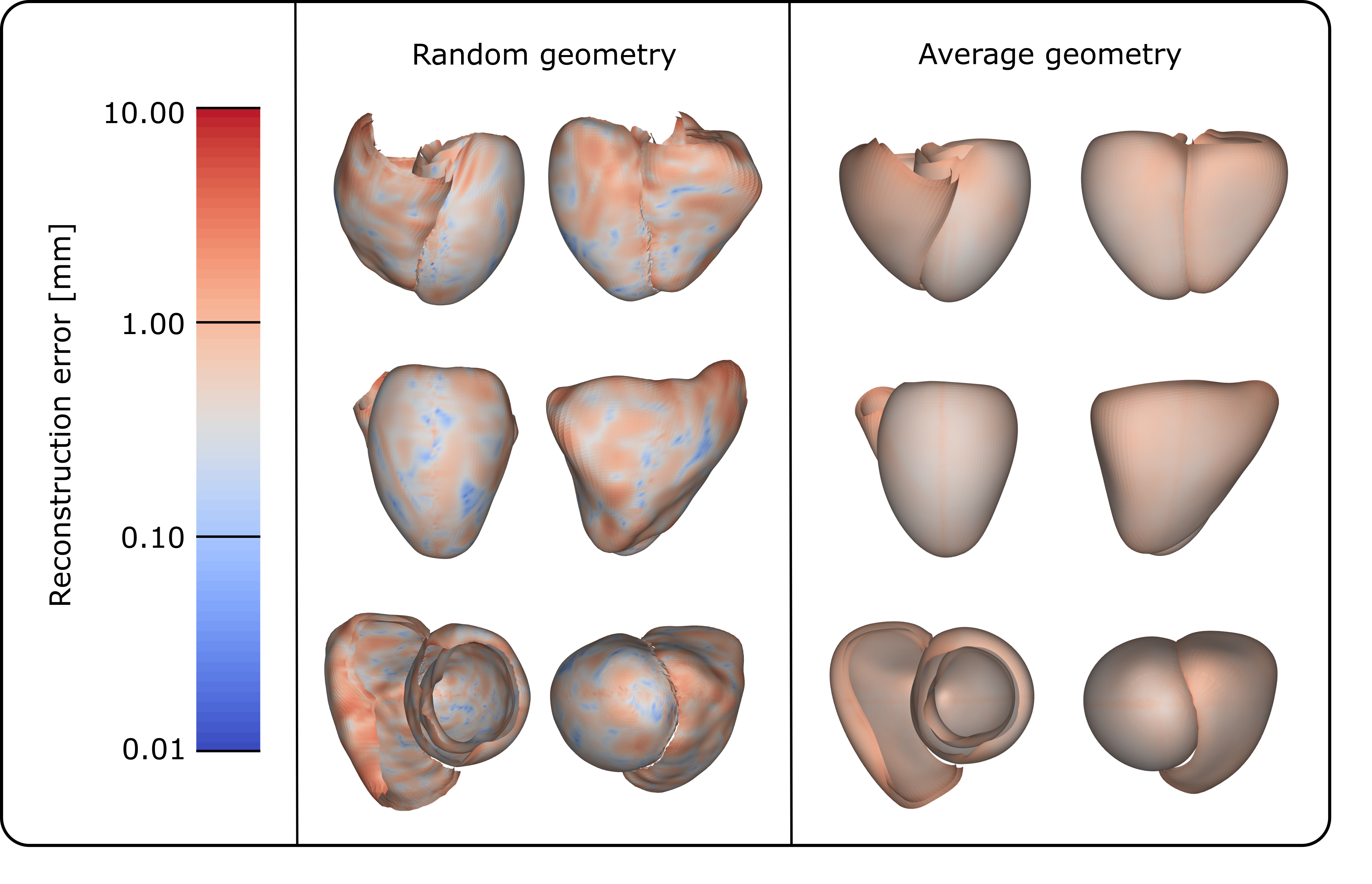}
	\caption{Pointwise reconstruction error mapped to biventricular geometry. Left: subject-specific pointwise reconstruction errors mapped onto the subject-specific reconstructed geometry, for a random subject. Right: average pointwise reconstruction errors mapped onto the average geometry of the complete dataset. Views from left-to-right, top-to-bottom: anterior-posterior, posterior-anterior, left lateral, right lateral, basal-apical, apical-basal.}
	\label{fig:reconstruction_error_full}
\end{figure}

Employing the LOOCV method, the reconstruction error was calculated for each leave-out geometry. The average reconstruction error obtained in this experiment was equal to 0.61~\SI{}{\milli \meter}, only slightly higher than the spatial resolution of the CT scans used for building the training set of the SSM. Figure~\ref{fig:reconstruction_error_full} shows the pointwise reconstruction errors mapped onto the mesh for a random subject, showing that the largest errors (8.83~\SI{}{\milli\meter}) are located at the RV apex, followed by the ventricular base, particularly at the junction between the LV and RV. To determine if this trend is consistent across geometries, we mapped the average pointwise reconstruction error, which ranges from 0.36 to 3.29~\SI{}{\milli\meter} across the different points, from the entire LOOCV experiment onto the mean biventricular geometry, as shown in Figure~\ref{fig:reconstruction_error_full}. This similarly highlights the basal region of both ventricles as the area with the highest error, along with a smaller area at the apex, near the junction between the LV and RV, but also shows a relatively uniform error distribution without excessively error-prone regions.

\subsection{Influence of demographics}

\begin{figure}[!t]
	\centering
	\includegraphics[width=0.48\textwidth]{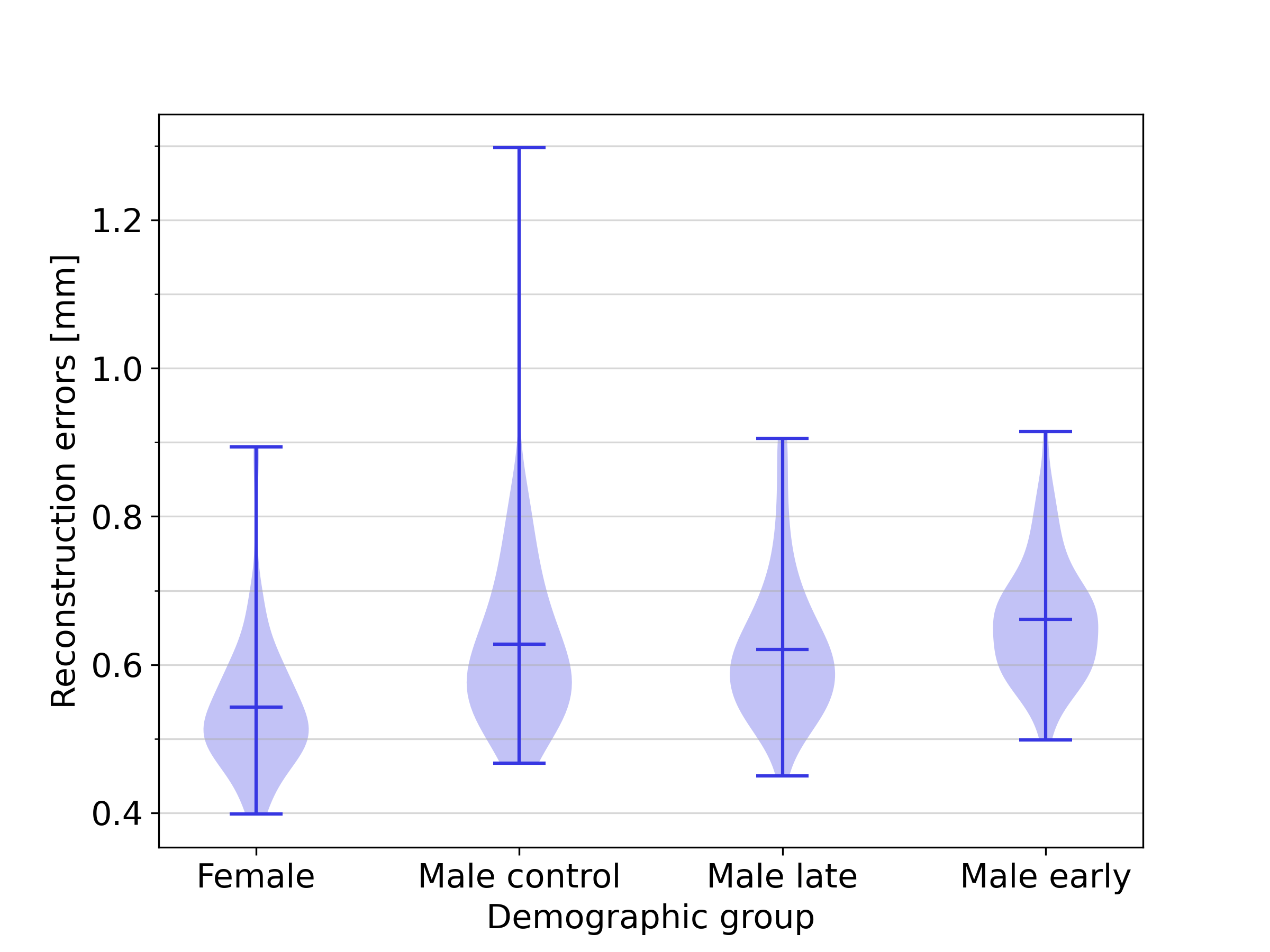}
	\caption{Reconstruction errors in the leave-one-out cross-validation experiment incorporating all subjects, displayed per demographic group. The means, extrema, and distributions are shown using violinplots.}
	\label{fig:loocv}
\end{figure}

Figure~\ref{fig:loocv} illustrates the distribution of the reconstruction errors for the full SSM across the demographic groups included in this study. The mean reconstruction error for the female group was 0.54 \SI{}{\milli\meter}, while those for the male control group, male early-onset athletes, and male late-onset athletes were 0.63 \SI{}{\milli\meter}, 0.66 \SI{}{\milli\meter} and 0.62 \SI{}{\milli\meter}, respectively. The results show that the female group can be approximated with a slightly smaller reconstruction error. No substantial differences in reconstruction accuracy were obtained for the other groups, indicating that the full SSM is equally effective in representing biventricular geometries across the three demographic groups of the male population. An additional analysis in \ref{app-1} shows that this effect is not caused by the slight demographic imbalance in the dataset, and \ref{app0} shows that the smaller reconstruction errors associated with female harts can largely, but not entirely be explained by heart size factors.

\subsection{Impact of dataset size}
\label{dataset_size}

\begin{figure}[!t]
	\centering
	\includegraphics[width=0.48\textwidth]{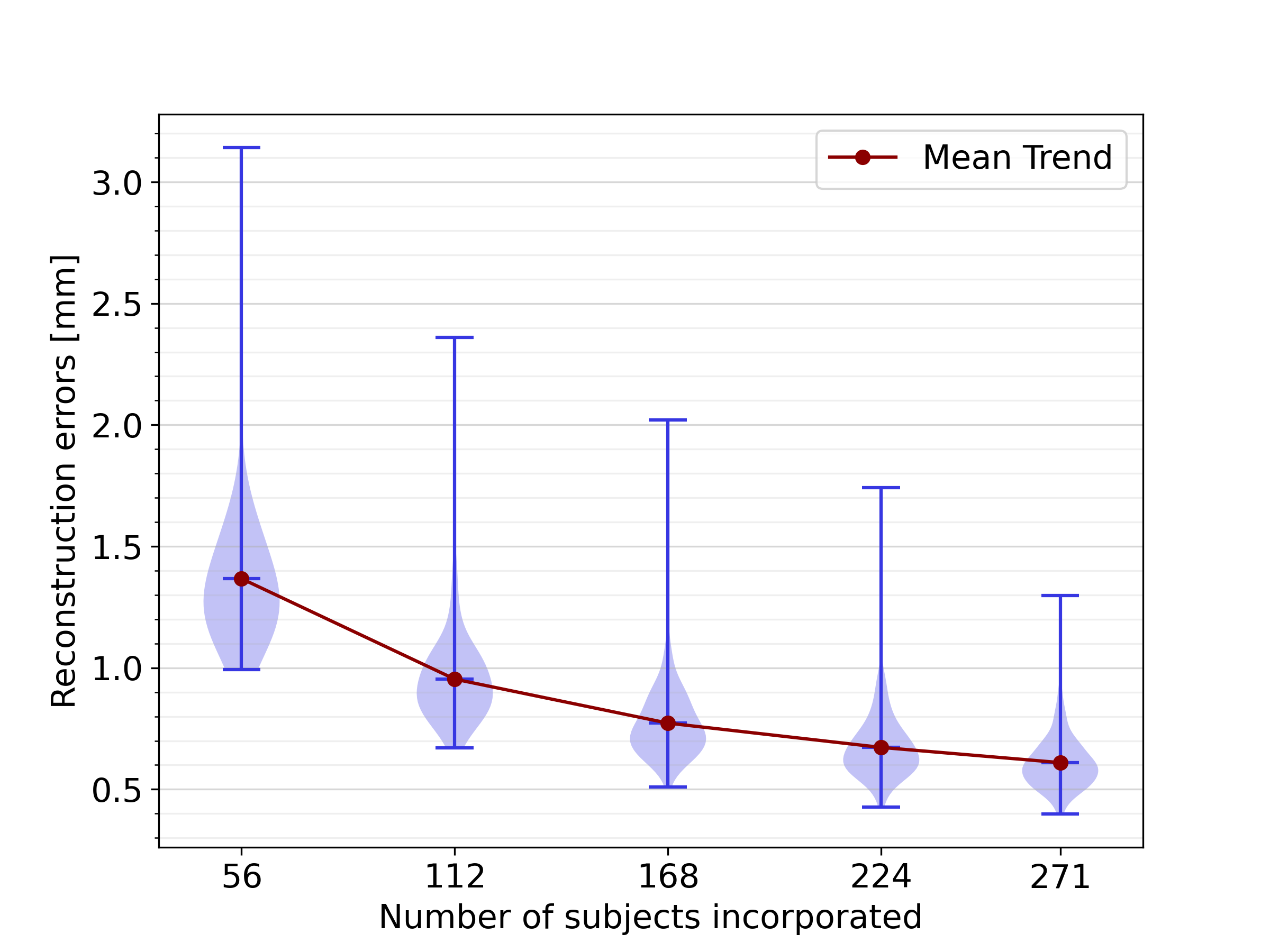}
	\caption{Decrease of reconstruction errors in the leave-one-out cross-validation experiments with increasing numbers of subjects for SSM construction. The means, extrema, and distributions are shown using violinplots.}
	\label{fig:loocv_subjects}
\end{figure}

Figure~\ref{fig:loocv_subjects} illustrates the results of LOOCV experiments conducted on data subsets of different sizes. The trend demonstrates that increasing the dataset size improves the SSM accuracy in representing healthy biventricular anatomies. Specifically, as the dataset size grows, the mean reconstruction error decreases significantly, from 1.37~\SI{}{\milli\meter} with 56 subjects to 0.61~\SI{}{\milli\meter} with all 271 subjects, indicating that further dataset expansion could potentially enhance the model performance even more.

\subsection{Impact of sex-related anatomical differences}
\label{PCdistributions}

A qualitative comparison of the first three SSM modes between the biased and unbiased SSMs revealed that the interpretation of these PCs was consistent, indicating a similar anatomical variation was captured with male-only, female-only and mixed-sex datasets.

\begin{figure}[!t]
	\centering
	\includegraphics[width=0.48\textwidth]{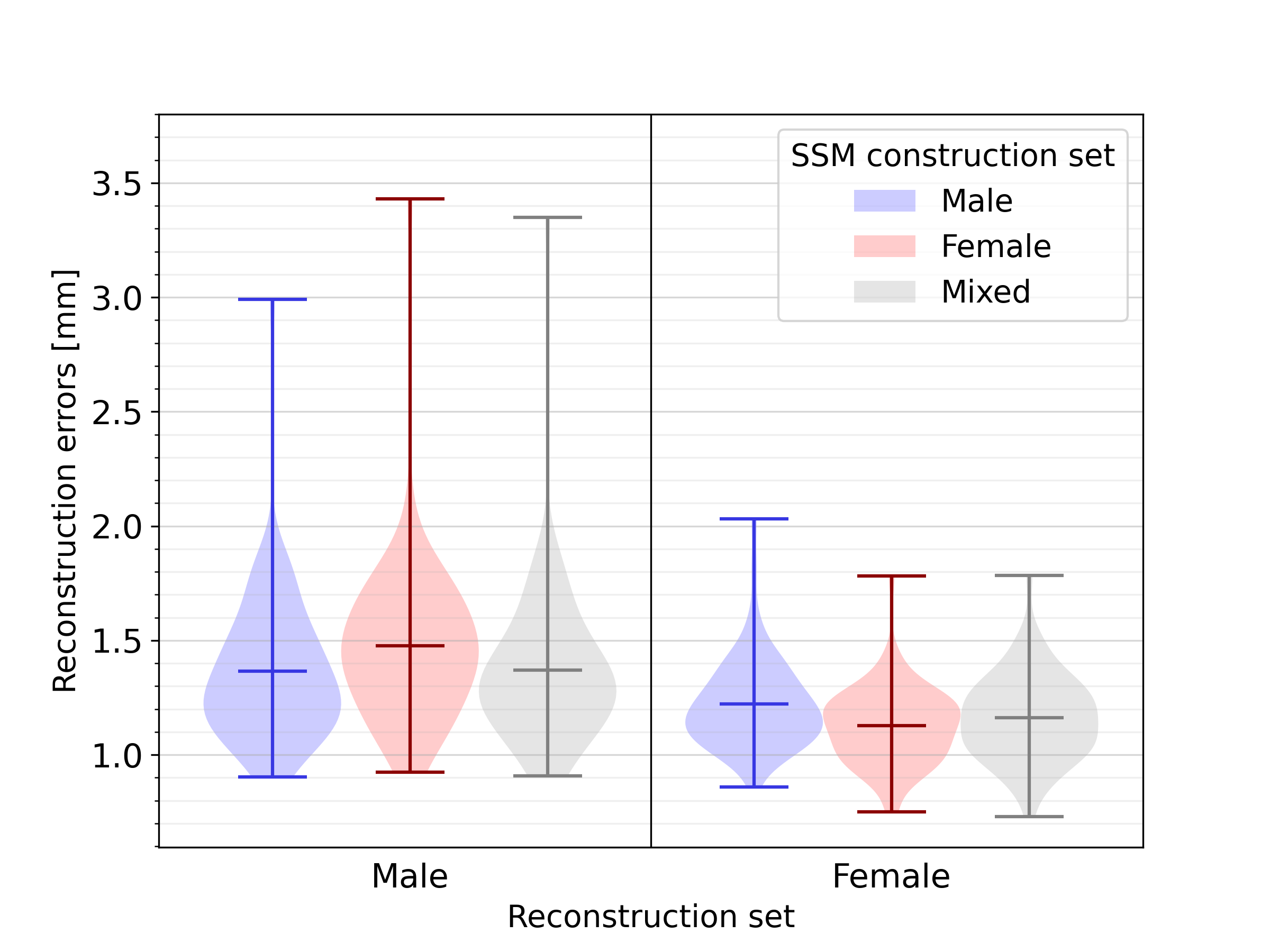}
	\caption{Reconstruction errors for male and female subjects in a series of SSMs constructed using a male, female or mixed dataset. The means, extrema and distributions are shown using violinplots.}
	\label{fig:malefemale}
\end{figure}

Next, a quantitative analysis of the impact of sex-related anatomical differences on the SSM was conducted. The results of reconstructing male and female geometries on two sex-biased SSMs, and one unbiased SSM, are shown in Figure~\ref{fig:malefemale}. For the reconstruction of male geometries, the reconstruction error was highest, namely 1.48~\SI{}{\milli\meter}, when using an opposite-sex SSM. The reconstruction error when using a same- or mixed-sex SSM was lower, respectively 1.36 and 1.37~\SI{}{\milli\meter}. The same holds for female geometries, where an opposite-, same- and mixed-sex SSM leads to reconstruction errors equal to 1.22, 1.13, and 1.16~\SI{}{\milli\meter}, respectively. For an additional analysis of the influence of sex-related anatomical differences on SSM representativeness, independent of heart size, we refer to \ref{app0}.


\begin{figure}[!t]
	\centering
	\includegraphics[width=0.46\textwidth]{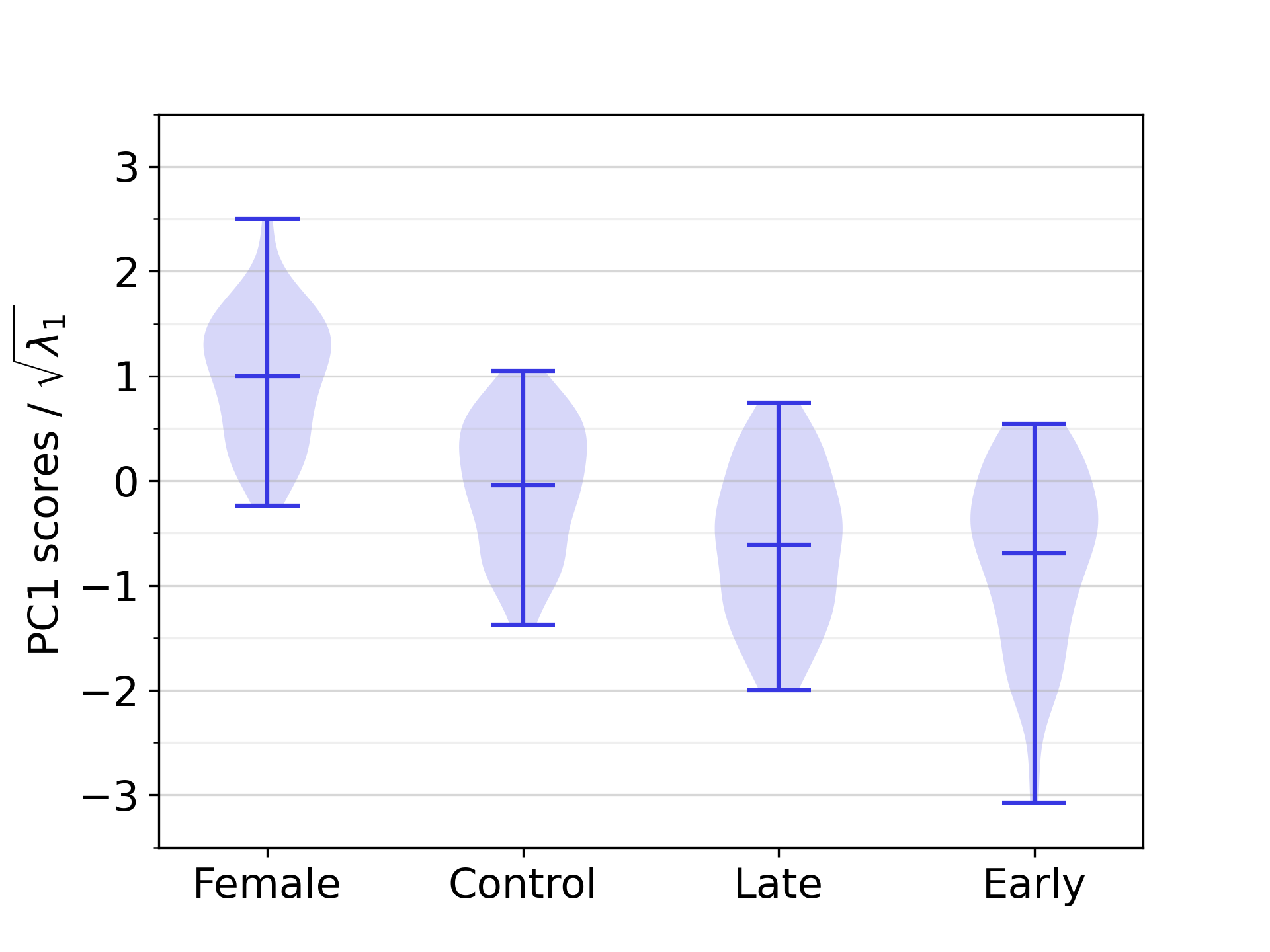}
	\caption{Distribution of scores for PC1 for different demographic subgroups. The scores are scaled by their standard deviation $\sqrt{\lambda_1}$.}
	\label{fig:PC1_subgroups}
\end{figure}

Regarding the distribution of PC scores across different demographic groups, we observed a statistically significant difference between the female group and the male control population in this dataset on the first PC (see Section \ref{stat_an} and Figure~\ref{fig:PC1_subgroups}). 

\subsection{Generation of synthetic geometries}

\begin{figure*}[!t]
	\centering
	\includegraphics[width=0.9\textwidth]{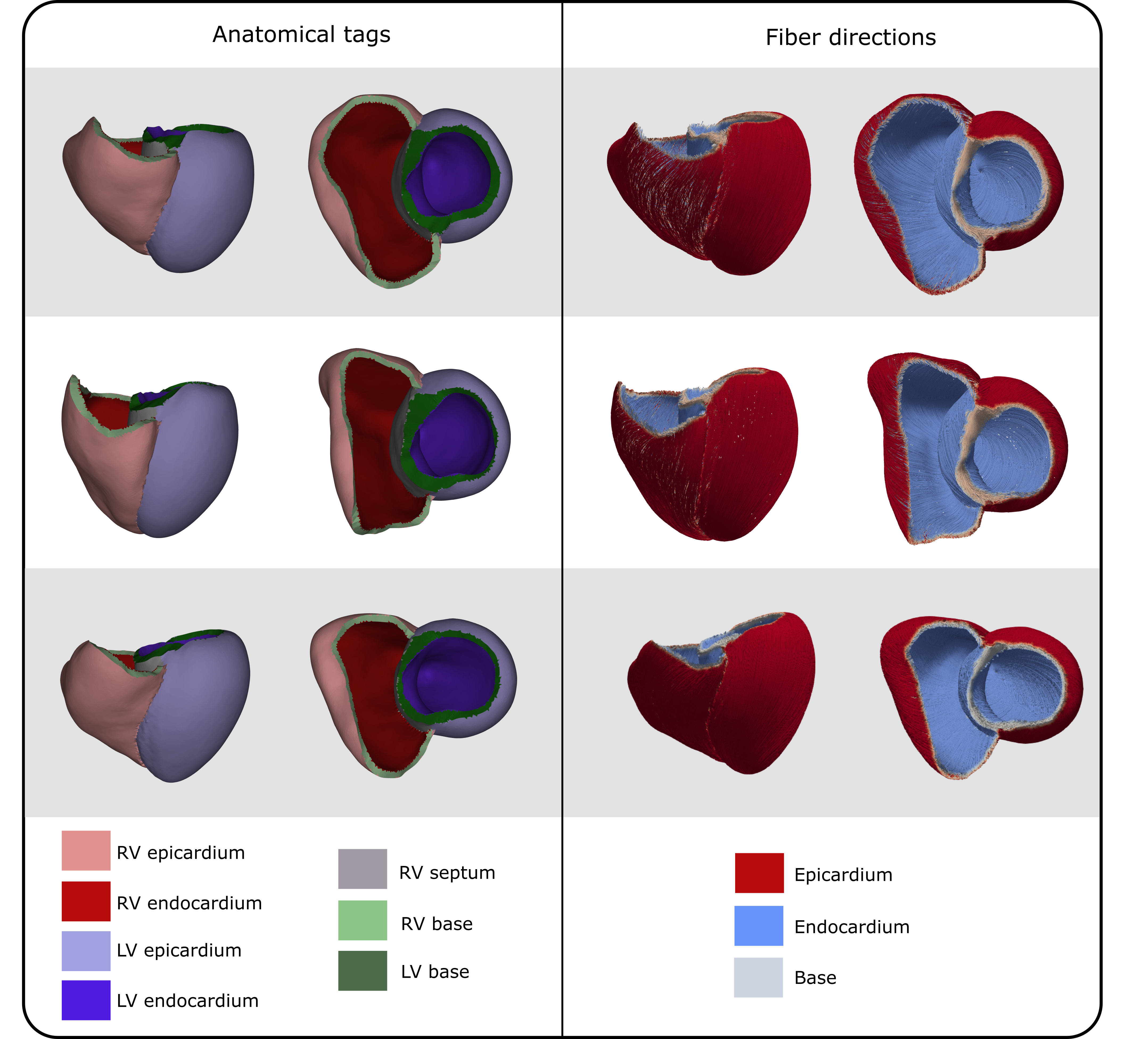}
	\caption{Synthetic geometries, obtained from sampling from 94 PCs of the full SSM. Each row shows one synthetic biventricular geometry, shown in anterior-posterior and basal-apical view, with a focus on anatomical tags (left) or fiber directions (right).}
	\label{fig:synthetic_tags_fibers}
\end{figure*}

Figure~\ref{fig:synthetic_tags_fibers} displays three example geometries of the synthetic cohort, generated by sampling from 94 PCs. The synthetic geometries were resampled to a mesh resolution of 0.9 \SI{}{\milli \meter}, and augmented with anatomical labels and fiber architectures.

\subsection{Statistical analyses}
\label{stat_an}

\begin{table}[!t]
\centering
\begin{tabular}{c c c c c}
\hline
\\[-1em]
\textbf{PC} & \textbf{Variable} & \textbf{R} & \textbf{p $\left[\times10^{-3}\right]$} \\ 
\\[-1em]
\hline
\\[-1em]
 1 & Ramp II & -0.41 & 0.00 \\ 
 1 & Ramp V6 & -0.38 & 0.00 \\ 
 1 & Samp II & 0.23 & 0.15 \\ 
 2 & Ramp II & -0.27 & 0.01 \\ 
 \\[-1em]
 \hline
 \\[-1em]
 1 & RRint & -0.50 & 0.00 \\ 
 1 & QTint & -0.34 & 0.00 \\ 
 1 & Tdur & -0.30 & 0.00 \\ 
 \\[-1em]
 \hline
 \\[-1em]
 1 & Axis & -0.24 & 0.10 \\ 
 \\[-1em]
 \hline
 \\[-1em]
 1 & Height & -0.62 & 0.00 \\ 
 1 & Weight & -0.37 & 0.00 \\ 
 1 & Age & 0.22 & 0.37 \\ 
 1 & BSA & -0.55 & 0.00 \\ 
 2 & Height & -0.24 & 0.08 \\ 
 2 & BMI & 0.36 & 0.00 \\ 
 \\[-1em]
 \hline
\end{tabular}
\caption{Significant correlations between principal component scores and numerical ECG-derived and demographic features, with their respective correlation coefficients. R indicates the correlation coefficient. The first group of features is amplitude-related, the second group interval-related, and the fourth group contains the demographic variables.}\label{table:corr_full}
\end{table}

The correlation analysis between PC scores and ECG-derived features revealed eight statistically significant correlations between biventricular anatomy and EP in our study population, of which the strongest and most significant one was the correlation between PC1 and RRint, with a correlation coefficient equal to -0.50. Furthermore, six statistically significant correlations were found between anatomical components and numerical demographic features, the strongest ones being the correlation between PC1 and height, followed by PC1 and BSA, with correlation coefficients equal to -0.62 and -0.55, respectively. The complete list of correlation coefficients is available in Table \ref{table:corr_full}. The difference between the scores of PC1 was significant between all demographic groups that were compared in statistical tests, except between the male early-onset athletes and the male late-onset athletes. For PC2 and PC3, no such statistically significant differences between demographic groups was identified. The relevant distributions are shown in \ref{app1}.

\subsection{Available material}

The biventricular SSM, including PCs, derived from the full population, is made publicly available on Zenodo \doi{10.5281/zenodo.14261122}, along with a synthetic cohort of 100 biventricular, volumetric meshes, including fiber orientation and anatomical tags for LV and RV endo- and epicardium, and LV septum. The volumetric meshes of the synthetic cohort are provided in VTK format. Code to generate surface meshes from the published SSMs, by sampling PC scores, is made available in the Github repository \url{https://github.com/LoreVanSantvliet/BiventricularSSM}. 

\section{Discussion}
\label{sec4}

This work reports on the development of an SSM for analyzing and generating healthy biventricular geometries. We propose the use of a lightweight UVC-based sampling method to achieve pointwise anatomical correspondence between different geometries. Our analysis of a population comprising male athletes, male healthy controls, and healthy female individuals reveals that biventricular size, elongation, and the relative positioning of the LV and RV are the three primary factors of variability, jointly explaining over 60\% of the shape variability in the dataset. Additionally, we identify 14 statistically significant correlations between biventricular anatomy, as quantified by the SSM, and other health-related variables, namely ECG-derived features and demographic information, in addition to 3 significant anatomical differences between demographic groups. Finally, we supply high-resolution volumetric biventricular meshes, designed for downstream applications such as electrophysiological simulations, thereby contributing valuable resources for future research and modeling efforts.

\subsection{UVC-based correspondence}

We establish the use of UVCs as a viable, lightweight alternative to traditional, nonlinear registration approaches typically used for shape registration, e.g. in \citep{bai_bi-ventricular_2015, mauger_right_2019, nagel_bi-atrial_2021}. This novel approach offers fine-grained control over the number and placement of points within the geometries. By employing UVCs, we streamline the SSM construction process and enable straightforward surface reconstruction from point cloud data. This flexibility and efficiency make UVC-based correspondence an attractive option for applications requiring fine-grained anatomical alignment, such as SSM construction.

\subsection{Evaluation of the full SSM}

The first PC of the SSM primarily represents biventricular size, consistent with prior findings that chamber size is a dominant source of anatomical variability in cardiac SSMs \citep{bai_bi-ventricular_2015, mauger_right_2019, rodero_linking_2021, nagel_bi-atrial_2021}. The second and third main PCs also echo patterns observed in previous studies: sphericity or elongation of chambers aligns with observations in \citep{bai_bi-ventricular_2015, mauger_right_2019, rodero_linking_2021}, while relative position and orientation of heart chambers are consistent with findings from \citep{rodero_linking_2021}. These similarities underscore the robustness of our SSM main components in capturing universal patterns of cardiac variability. The absence of unexpected components additionally provides an indirect validation of the suitability of UVC-based sampling in the SSM construction pipeline.

Our SSM captures 90\% of the anatomical variance in the population in 16 PCs, 95\% in 29 PCs, and 99\% in 94 PCs. These results closely align with previous studies on cardiac SSMs. For instance, Mauger et al. \citep{mauger_right_2019} reported that the first 50 PCs encoded 92.1\% of shape variance in a larger biventricular dataset of 4~329 subjects. Bai et al. \citep{bai_bi-ventricular_2015} showed that the first 8 PCs of their SSM, extracted from 1~093 subjects, explained over 90\% of the shape variance for the LV. Rodero et al. \citep{rodero_linking_2021} found that only 9 PCs were sufficient to capture 90\% of the variance of four-chamber shape variability, in a study incorporating 19 subjects. Nagel et al. \citep{nagel_bi-atrial_2021} found that 18 and 24 PCs accounted for 90\% and 95\% of the total shape variance, respectively, in a dataset of 47 geometries.  These differences between studies likely stem from various methodological factors, such as the spatial resolution of the tomographic images, the size and diversity of the dataset, the segmentation, alignment, and registration approaches, and the anatomical scope of the SSM. 

The average reconstruction error of 0.61 \SI{}{\milli\meter} in our LOOCV experiment is in line with values reported in the literature. A median reconstruction error of 1.56 \SI{}{\milli\meter} is reported for the atria in \citep{nagel_bi-atrial_2021}, while a mean reconstruction error around 0.15 \SI{}{\milli\meter} is reported for the ventricles in \citep{bai_bi-ventricular_2015}. However, this last, notably small reconstruction error was calculated on the set of geometries used for the SSM training, limiting the comparability to our metric. In \citep{mauger_right_2019}, the authors reported fitting errors ranging from 1.2 to 1.5 \SI{}{\milli\meter} for ventricular models, computed by point-to-surface distances for the registration step specifically, thus not aligning with our approach. Due to the absence of a standardized way of measuring SSM performance, comparing SSM quality across works in the literature remains a challenging task. However, by comparing our LOOCV metric with the identically computed one reported in \citep{nagel_bi-atrial_2021}, we conclude that our SSM offers competitive representativeness for biventricular geometries in a healthy population. 

\subsection{Biventricular anatomy in demographic groups}
\label{anat-dem}

Our study population includes both male and female subjects, enabling the investigation of sex-related differences in our cardiac SSM. Previously, \citep{salton_gender_2002, linde_interaction_2018, pfaffenberger_size_2013} reported that female ventricles are significantly smaller than male ones, based on various measurements of both the LV and RV, such as ventricular mass, volume, diameters, and lengths. Our study replicates such findings by showing that on average, female ventricles are smaller in size (as captured by PC1) than those of male individuals in the control population, as shown in Figure~\ref{fig:PC1_subgroups}. This finding is statistically significant, as explained in Section \ref{stat_an}.

However, St. Pierre et al. point out that the female heart is not just a small version of the male heart, i.e. the differences are more complex than those obtained from simple geometric scaling \citep{st_pierre_sex_2022}. More specifically, in \citep{st_pierre_sex_2022}, the authors indicated a non-isometric scaling of myocardial wall thickness with ventricular mass, for both the LV and RV. While the first PC of our SSM appears to capture size differences, it remains unclear whether this represents simple geometric scaling or incorporates other structural variations. Further investigation is needed to explore these aspects in more detail. While patient-specific LV myocardial thickness is preserved in our anatomical generation pipeline, variations in the RV myocardial thickness are not captured, but fixed, as explained in Section \ref{uvcs}. 

Statistically significant differences were also observed in PC1 between the healthy control group, and both early- and late-onset athletes, indicating that both groups of male athletes have significantly larger ventricles compared to male subjects in the healthy control group. This finding is consistent with existing literature on athletic hearts. Morganroth et al. reported an increased LV end-diastolic volume and mass for athletes engaging in isotonic exercise, compared to healthy controls \citep{morganroth_comparative_1975}. Similarly, Oxborough et al. found that RV chamber dimensions in athletes often exceed normal ranges, particularly towards the large end of the spectrum \citep{oxborough_right_2012}.

\subsection{Sex-related anatomical differences}

There is a growing interest in sex-related differences in cardiac anatomy and function, as these differences are crucial for enhancing cardiovascular care and ensuring equitable treatment across sexes \citep{vogel_lancet_2021, reue_illuminating_2022, st_pierre_sex_2022, burrowes_sex_2023}. We performed a variety of experiments to assess the impact of sex on the performance and representativeness of the SSM. As described in Section \ref{anat-dem}, one key finding was the significant difference in biventricular size that we observed between female subjects of our study population, and male subjects belonging to the healthy control group. This sex-related difference is consistent with findings by Burns et al. \citep{burns_genetic_2024}.

Our results highlight the importance of using unbiased datasets in terms of sex, in order to be representative of both male and female anatomy. For both sexes, an unbiased SSM performed a better reconstruction than a biased SSM built from geometries of the opposite sex. The difference in reconstruction error when using a biased SSM built from geometries from the same sex or using an unbiased SSM, was minor in both cases. Notably, we found the average reconstruction error for female subjects to be consistently smaller than for male subjects, regardless of whether the SSM used for reconstruction is biased or not, probably due to the overall smaller heart size.

\subsection{Impact of dataset size}

We investigated the reconstruction error in a series of LOOCV experiments varying the SSM training set size. The decreasing trend in reconstruction error when increasing the training set size is in line with the findings of Bai et al. \citep{bai_bi-ventricular_2015}, who showed a substantial reduction in reconstruction error for ventricular geometries with dataset sizes up to 400-600 subjects, when only the first 100 PCs were considered for reconstruction. Our findings, combined with those of Bai et al., suggest that a straightforward improvement to our model would be to include an even larger number of healthy subjects. This could not only further reduce reconstruction error, but also increase the statistical power when correlating PC scores to other health-related variables, possibly enabling the identification of additional, significant sex-related differences between PCs.

\subsection{Synthetic cohort and CDTs}

Leveraging the diversity captured by our biventricular SSM, we created a synthetic cohort of anatomically detailed biventricular meshes. The geometries in this cohort were each annotated with anatomical labels and fiber orientations. This cohort, and possibly even more synthetic geometries derived from our SSM, can be used for large-scale simulation studies. In addition to the fundamental insights that large-scale simulation studies can provide on simulation technology, for example by clarifying their parameter sensitivities and uncertainties \citep{zappon_quantifying_2024}, they can also be used to create simulation databases \citep{gillette_medalcare-xl_2023}. Simulation databases, in turn, are a useful resource for training data-driven calibration models to perform automated functional personalization (calibration) of patient-level CDTs. A simulation database suitable for training advanced, anatomy-aware, data-driven calibration models, such as DL models, needs to be large and diverse. Both of these requirements are made more accessible by the release of our SSM, which offers anatomical diversity without requiring researchers to access or process sensitive health data, and which offers the ability to generate a virtually unlimited amount of synthetic, yet realistic heart shapes.

Additionally, the synthetic cohort itself serves as an important building block toward population-level CDTs. Such CDTs, also called virtual cohorts, have application potential in virtual clinical trials \citep{corral-acero_digital_2020, niederer_creation_2020}. We envision the use of our anatomical cohort of ready-to-use biventricular meshes, representative of a healthy population, to create a healthy virtual control population for such trials.

\subsection{Correlations}

In an exploratory analysis, we identify several statistically significant correlations between PC scores and other health-related variables in our study population. For reasons related to statistical power, we limit the number of statistical tests in two ways. First, we only consider the first three PCs. Secondly, for amplitude-related features, which could in principle be derived from any lead, we select only two leads: lead II for its general clinical relevance \citep{meek_introduction_2002} and lead V6 for its connection to hypertrophy \citep{romhilt_critical_1969}. 

Among all amplitude-related ECG features, only R and S wave amplitude were found to be significantly correlated with anatomical features. More specifically, we observe a negative correlation between PC1 and Ramp of both lead II and lead V6, indicating that larger ventricles are associated with larger R wave amplitudes. Additionally, a positive correlation between PC1 and S wave amplitude of lead II, combined with the fact that S wave amplitudes are negative, suggests that larger hearts have larger S waves. Furthermore, a negative correlation between PC2 and Ramp of lead II is detected, suggesting that more elongated ventricles are associated with larger R waves.

The positive correlation between heart size and R wave amplitude is consistent with simulation-based findings: Zappon et al. \citep{zappon_quantifying_2024} observed an increase in simulated R wave amplitude in lead II and lead V6 when increasing heart size by geometric scaling, although their analysis was based on a single geometry. Moreover, the increase of R wave amplitude in lead V6 is consistent with diagnostic criteria for ventricular hypertrophy \citep{romhilt_critical_1969}.

For interval-related features, a significant positive correlation was found between PC1 and RRint, QTint, and Tdur. This indicates that larger hearts tend to have longer RR and QT intervals, as well as longer T waves. Notably, no significant correlation was found between PC1 and QTcint, where QT interval length is corrected for the influence of the overall RR interval duration.

The positive correlation between heart size and RR interval is consistent with prior research: LV mass index has been related to the RR interval through correlation with the heart rate \citep{saba_gender_2001}. Additionally, Burns et al. also detected a significant negative correlation between heart size and heart rate \citep{burns_genetic_2024}. In electromechanical simulations, Rodero et al. \citep{rodero_linking_2021} observed a positive correlation between the second mode of variation in their SSM, encoding heart size, and the duration of systole, closely related to the QT interval. This supports our findings that larger hearts tend to have longer QT intervals. We did not, however, find studies validating the correlation between PC1 and T wave duration. 

Finally, we found a negative correlation between PC1 and the electrical axis, either indicating a relation between heart size and orientation, or revealing a bias introduced by heart size when calculating the electrical axis based on ECG signals.

Regarding demographic features, the first PC, which encodes ventricular size, is negatively correlated with height, weight, and BSA. This suggests that larger values for each of these demographic variables are associated with larger ventricles. Additionally, a negative correlation is present between PC2 and height, indicating that taller subjects tend to have more elongated ventricles. The positive correlation between PC2 and BMI indicates that individuals with a higher BMI tend to have more round ventricles. Finally, there is a positive correlation between age and PC1, implying that older subjects typically have smaller hearts in our study population.

The significant correlation between a principal component encoding ventricular size and height is confirmed in \citep{burns_genetic_2024}. Both the correlation between ventricle elongation and BMI, and between heart size and age are consistent with findings of Burns et al. \citep{burns_genetic_2024} as well. Moreover, the correlations of height, weight and BSA with PC1 aligns with the work of Hense et al. \citep{hense_associations_1998}, who identified significant positive correlations between LV mass on the one hand, and weight, height, and BMI on the other hand. However, LV mass and BSA correlation were not investigated in this study. Finally, in a series of multivariate linear regression analyses, Pfaffenberger et al. \citep{pfaffenberger_size_2013} examined the influence of demographic factors on a range of variables related to ventricular size (LV end-diastolic and end-systolic diameter, LV end-diastolic and end-systolic volume, RV end-diastolic diameter, RV end-diastolic area). They found significant effects of height, weight, and BSA on most of these variables, consistent with our results.

\subsection{Applications in patient-level CDTs}

Our automated workflow for anatomical twinning is directly applicable for generating patient-level CDTs. Beyond this methodological applicability, the insights that are drawn from our statistical analysis of the relationship between ventricular anatomy and electrophysiology, are more broadly useful for CDT research. As an example, we propose their use for establishing which anatomical factors are most important to capture in a CDT. For example, statistical correlations revealed in this work, presented in Table~\ref{table:corr_full}, confirm that especially heart size is very important to model accurately in CDTs, as it has a significant effect on many different aspects of clinical ECGs.

A second possible application of the statistical insights is in validation of CDT simulation technology. Validation of CDT technology is highly challenging \citep{sel_building_2024}. How can we confirm that a set of PDEs adequately reflects the physiological behavior of the cardiovascular (sub)system under investigation for a given application? Validation approaches for cardiac digital twins in healthcare rely on different types of evidence \citep{pathmanathan_validation_2018}. The first category consists of evidence related to the mathematical equations employed, including modeling assumptions and parameters. The second category, calibration evidence, involves demonstrating alignment between the model's simulated or predicted results and the clinical data used for calibration. Finally, validation evidence pertains to the model's ability to accurately replicate real-world data that was not used at any stage of model construction.

Linking anatomical analysis results to clinical, ECG-derived features and demographic information, provides a foundation for validating the simulation technology underlying CDTs of cardiac EP. Statistically significant correlations between anatomical factors and clinical ECG-derived features should also appear between these same geometrical factors and simulation-based ECG features. Such findings can thus be used to verify whether existing modeling equations display the behavior observed in clinical reality. These correlations can serve as calibration or validation evidence, depending on whether the validation would be conducted on subjects used for building the SSM or on independent ones. As such, we propose the use of the SSM in the context of CDT validation as a possible future extension of our work.

\subsection{Limitations and future extensions}

For male subjects, data was derived from the Master@Heart study, a prospective cohort trial including athletes and healthy control subjects \citep{de_bosscher_endurance_2021}. For female subjects, on the other hand, data was obtained from a clinical referral of the subjects to a cardiac CT. These subjects can still be considered healthy, since they are not diagnosed with any cardiovascular disease, and subjects that presented with more than three cardiovascular risk factors were excluded. However, their healthy status might still not match that of subjects in the Master@Heart study, with more selective exclusion criteria.

Our SSM covers biventricular cardiac anatomy below the cardiac valves, excluding the non-conductive cardiac skeleton. Given the electrophysiological focus of our work, this non-conductive tissue is less important to incorporate into our SSM. However, for enabling the use of the provided meshes for mechanical or hemodynamical modeling, valvular anatomy should be included. In this case, an adapted UVC system that includes cardiac valves, such as the one provided in \citep{pankewitz_universal_2024}, could be valuable for extending our UVC-based approach to establish correspondence between geometries to valvular structures. If the extended UVCs are ordered in a sufficiently regular fashion in Cartesian space to enable subsequent triangulation, as covered in section \ref{surface_construction}, an extension of the SSM towards additional anatomical structures would then be straightforward. However, incorporating additional structures, and hence possibly introducing additional anatomical variability is expected to further increase data requirements for obtaining a representative SSM.


Additionally to the ventricles, our SSM could be extended towards the atrial structures as well, leading to a four-chamber SSM. Given the importance of the atria in establishing P wave morphology, this would be an interesting extension, allowing further investigation of relationships between cardiac anatomy and electrophysiology. Given that atria are structurally highly complex \citep{nagel_bi-atrial_2021}, arguably even more so that ventricles, it is expected that a larger dataset would be more suitable for this extension. To include the atria, the correspondence establishment could be extended by making use of universal atrial coordinates \citep{roney_universal_2019, zappon_efficient_2025} in addition to UVCs. Similarly to before, the ordering of these coordinates needs to be sufficiently regular as to allow subsequent triangulation, possibly with the aid of Laplacian smoothing. A suitable sampling scheme over the different atrial coordinates needs to be devised, leading to an additional number of points per mesh, thus an additional number of input features for the PCA.

The results described in Section~\ref{dataset_size} reveal that increasing the dataset would likely lead to further improvements of the SSM. The UK Biobank, used in \citep{mauger_right_2019, burns_genetic_2024}, is a suitable resource for increasing the dataset size, although the imaging resolution of these cardiac MR images is lower than the resolution of CT scans used in this study.

\section{Conclusion}
\label{sec5}

This study reports on a highly automated workflow for the construction of an SSM of biventricular anatomy. The training set was built upon high-resolution cardiac CT scans of 271 healthy subjects, including athletes. A novel, lightweight UVC-based approach for establishing correspondence between geometries was introduced. The resulting SSM was used in various downstream analyses revealing quantitative relationships between ventricular anatomy and ECG-derived features, as well as demographic features. Additionally, a virtual cohort of 100 volumetric, biventricular meshes, designed for use in electrophysiological simulations, was extracted from the SSM and published.

\section*{Ethical approval}

Ethical approval for the reuse of the Master@Heart dataset in this study was granted by the Ethics Committee Research UZ/KU Leuven on March 29th 2024, under reference number S61336. Ethical approval for the use of the clinical data of the female subjects in this study was granted by the Ethics Committee Research UZ/KU Leuven on March 29th 2024, under reference number S68896.

\section*{Funding sources}

This research is funded by a PhD fellowship fundamental research from the Research Foundation - Flanders, Belgium (FWO) for L. Van Santvliet (FWO project number 1107725N); the Flanders AI Research Program, Belgium; HORIZON-HLTH-2022-IND-13: `Privacy compliant health data as a service for AI development (PHASE IV AI)', European Union (grant agreement no. 101095384); a research grant from the Research Foundation - Flanders, Belgium (FWO) for M. De Vos (FWO project number G046925N); a research grant from the Research Foundation - Flanders, Belgium (FWO) for R. Willems (FWO project number T003717N). This research received support from the European Union's Horizon 2020 research and innovation program under the Marie Sk\l{}odowska-Curie grant TwinCare-AF (grant agreement no. 101148636) for E. Zappon, who also acknowledges her membership to INdAM GNCS - Gruppo Nazionale per il Calcolo Scientifico (National Group for Scientific Computing, Italy). Further financial support was received from the Austrian Research Promotion Agency (FFG) within the InstaTwin project, grant FO999891133, and from the Austrian Science Fund (FWF), grant 10.55776/PAT174842.

\begin{figure}[hbt!]
\centering
\includegraphics[width = 0.4\textwidth]{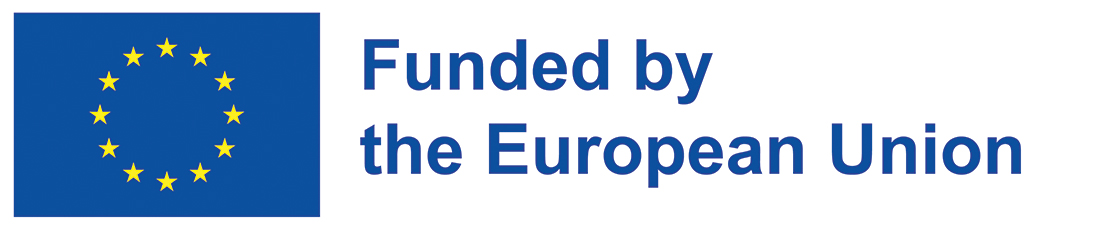}
\end{figure}


\appendix

\section{Evaluation of UVC-based sampling}\label{app_sampling}

We studied the impact of the number of mesh points, termed the mesh size, on the sampling error. Starting from the original \SI{0.9}{\milli\meter} biventricular meshes, constructed as explained in Section~\ref{uvcs}, we resampled all 271 meshes of the complete dataset using a variety of sampling schemes, displayed in Table~\ref{table:sampling}. By using a constant scaling factor for adapting each of the three meshing parameters (the number of $z$ values,  the number of $\varphi$ values for the LV and the number of $\varphi$ values for the RV), the element shapes are kept constant over all meshing schemes. An identical apical coarsening procedure to the one described in Section~\ref{surface_construction} was followed for all sampling schemes. Sampling errors were defined as point-to-surface distances of resampled mesh points to the original mesh, averaged over all mesh points. They were calculated using the \texttt{vtkDistancePolyDataFilter} (Kitware, Clifton Park, New York, USA).

\begin{table}[!t]
\centering
\begin{tabular}{c c c c c}
\hline
\textbf{ } & \textbf{\# $z$ values} & \textbf{\# $\varphi_{LV}$ values} & \textbf{\# $\varphi_{RV}$ values} & \textbf{Size}\\
\hline
 1 & $20$ & $40$ & $20$ & $2084$ \\
 2 & $20\times2=40$ & $40\times2=80$ & $20\times2=40$ & $8064$ \\
 3 & $20\times3=60$ & $40\times3=120$ & $20\times3=60$ & $17944$ \\
 4 & $20\times4=80$ & $40\times4=160$ & $20\times4=80$ & $31724$ \\
 5 & $20\times5=100$ & $40\times5=200$ & $20\times5=100$ & $49404$ \\
 \hline
\end{tabular}
\caption{Summary of the five sampling schemes, showing the number of $z$ values, the number of $\varphi$ values for the LV and RV, and the mesh size. The number of mesh points is dictated by the number of $z$ and $\varphi$ values selected, and apical coarsening. Sampling scheme 4 was used throughout the paper.}\label{table:sampling}
\end{table}

\begin{figure}[!t]
	\centering
	\includegraphics[width=0.48\textwidth]{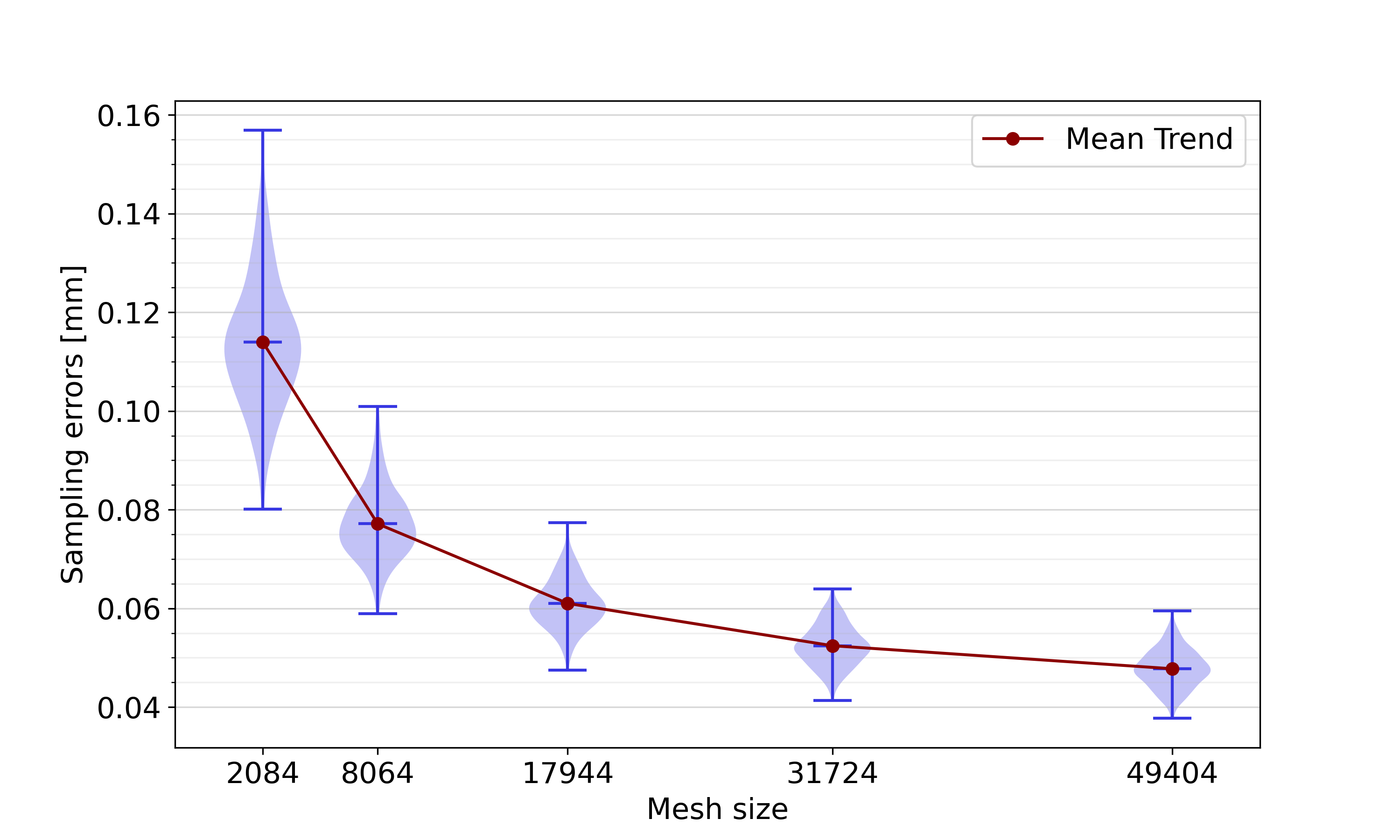}
	\caption{Decrease of sampling errors with increasing mesh sizes for UVC-based resampling. The means, extrema, and distributions are shown using violinplots.}
	\label{fig:sampling}
\end{figure}

Figure~\ref{fig:sampling} shows the distribution of the sampling errors of all meshes of the complete dataset for each of the five sampling schemes. A clear trend is visible, with lower sampling errors for increasing mesh sizes. This trend is expected: the more mesh points are included in a mesh, the better we expect it to represent shapes in their full complexity. However, it is also expected that at some point, the advantage of having a larger mesh size is not as pronounced anymore, i.e. the sampling error stagnates. This justifies our choice of sampling scheme~4.

\section{Biventricular anatomy in balanced demographic groups}\label{app-1}

\begin{figure}[!t]
	\centering
	\includegraphics[width=0.48\textwidth]{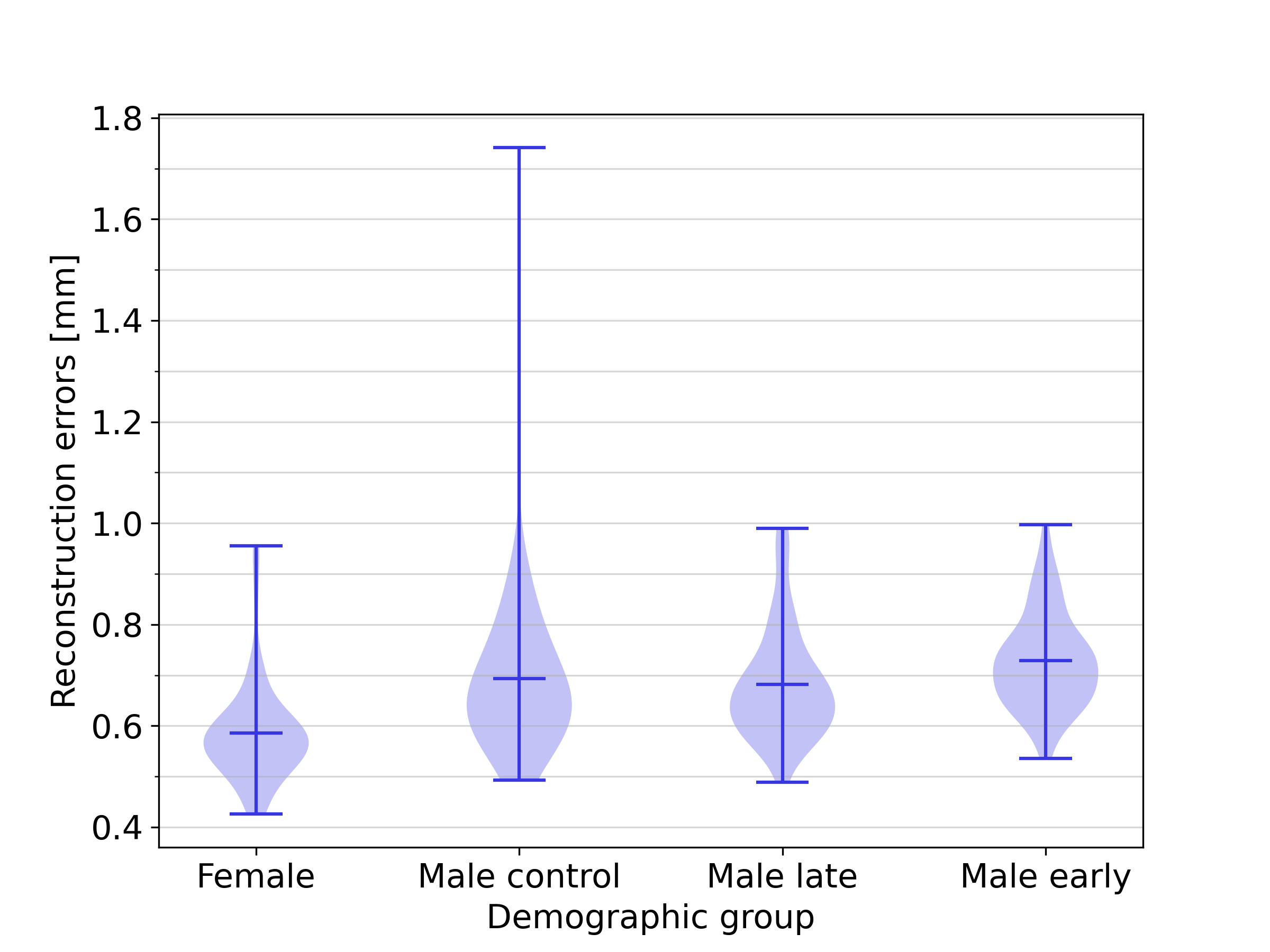}
	\caption{Reconstruction errors in the leave-one-out cross-validation experiment incorporating 56 subjects from each demographic group, displayed per demographic group. The means, extrema, and distributions are shown using violinplots.}
	\label{fig:loocv_balanced}
\end{figure}

The analysis reported in Section~\ref{anat-dem} was repeated using a balanced dataset containing 56 subjects from each of the four demographic groups instead of the slightly imbalanced complete dataset. The results of this analysis, displayed in Figure~\ref{fig:loocv_balanced} show an overall slightly increased reconstruction error due to the smaller overall dataset size: 0.59 \SI{}{\milli\meter} for the female group, and 0.69 \SI{}{\milli\meter}, 0.73 \SI{}{\milli\meter} and 0.68 \SI{}{\milli\meter} for the male control group, male early-onset athletes, and male late-onset athletes, respectively. Qualitatively, the differences in reconstruction error between the various demographic groups remain unchanged.

\section{Influence of shape factors independent of heart size}
\label{app0}

To reveal the effect of biventricular shape differences between demographic groups and between sexes, independent of heart size, additional analyses were performed. For these analyses, scaling was allowed in the Procrustes alignment, leading to SSMs where heart size was not encoded as one of the PCs. Apart from this change, an identical methodology to the one described in \ref{app-1} and Section \ref{eval} was used for the demographic groups and sexes, respectively.

\begin{figure}[!t]
	\centering
	\includegraphics[width=0.48\textwidth]{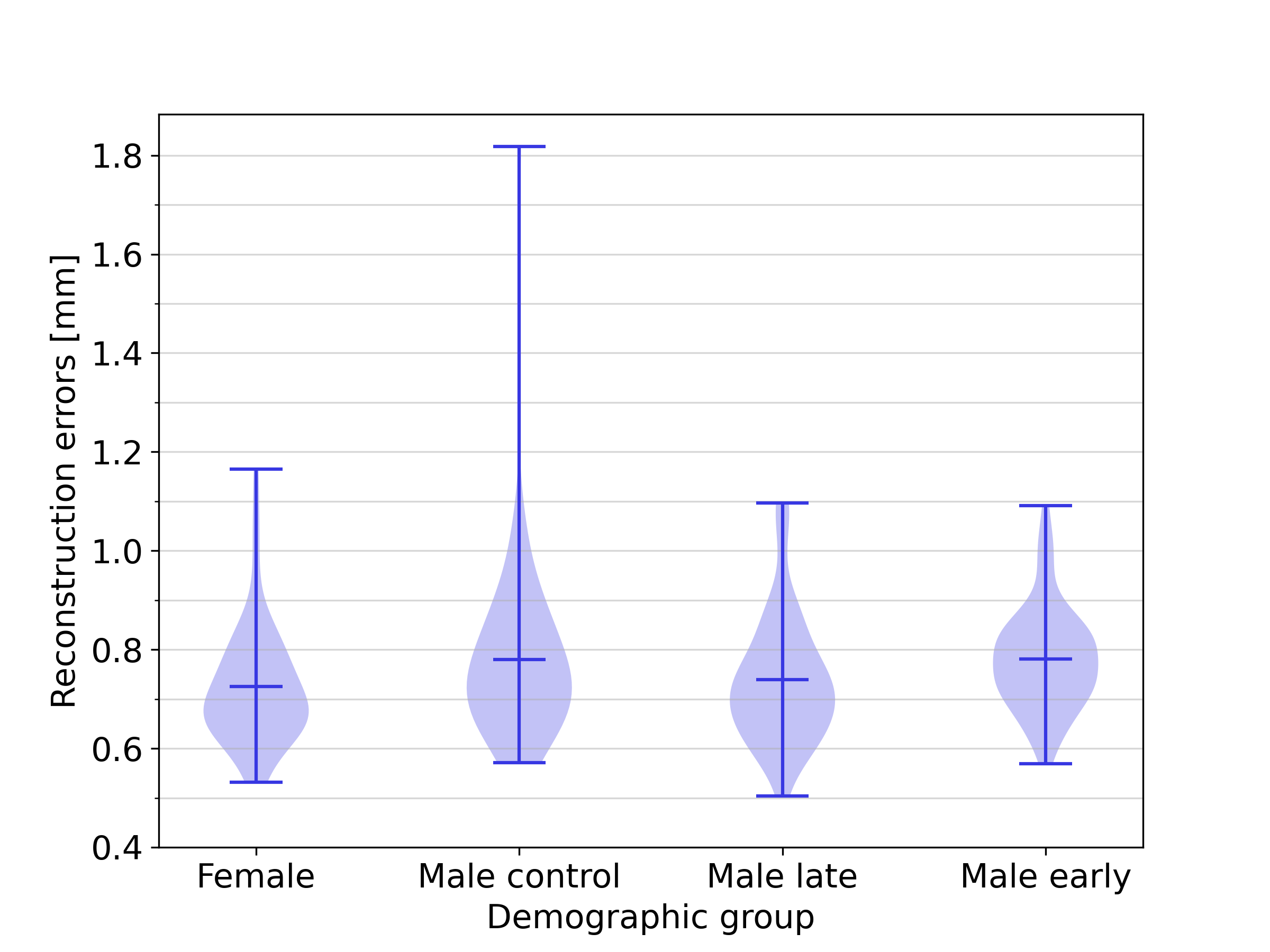}
	\caption{Heart size-independent reconstruction errors in the leave-one-out cross-validation experiment incorporating 56 subjects from each demographic group, displayed per demographic group. The means, extrema, and distributions are shown using violinplots.}
	\label{fig:loocv_nosize}
\end{figure}

Figure~\ref{fig:loocv_nosize} shows the reconstruction errors per demographic group. The female group was associated with the smallest reconstruction errors, although the difference was much smaller compared to Figure~\ref{fig:loocv_balanced}. Therefore, we conclude that the smaller reconstruction error for this demographic group can be largely but not entirely explained by heart size.

\begin{figure}[!t]
	\centering
	\includegraphics[width=0.48\textwidth]{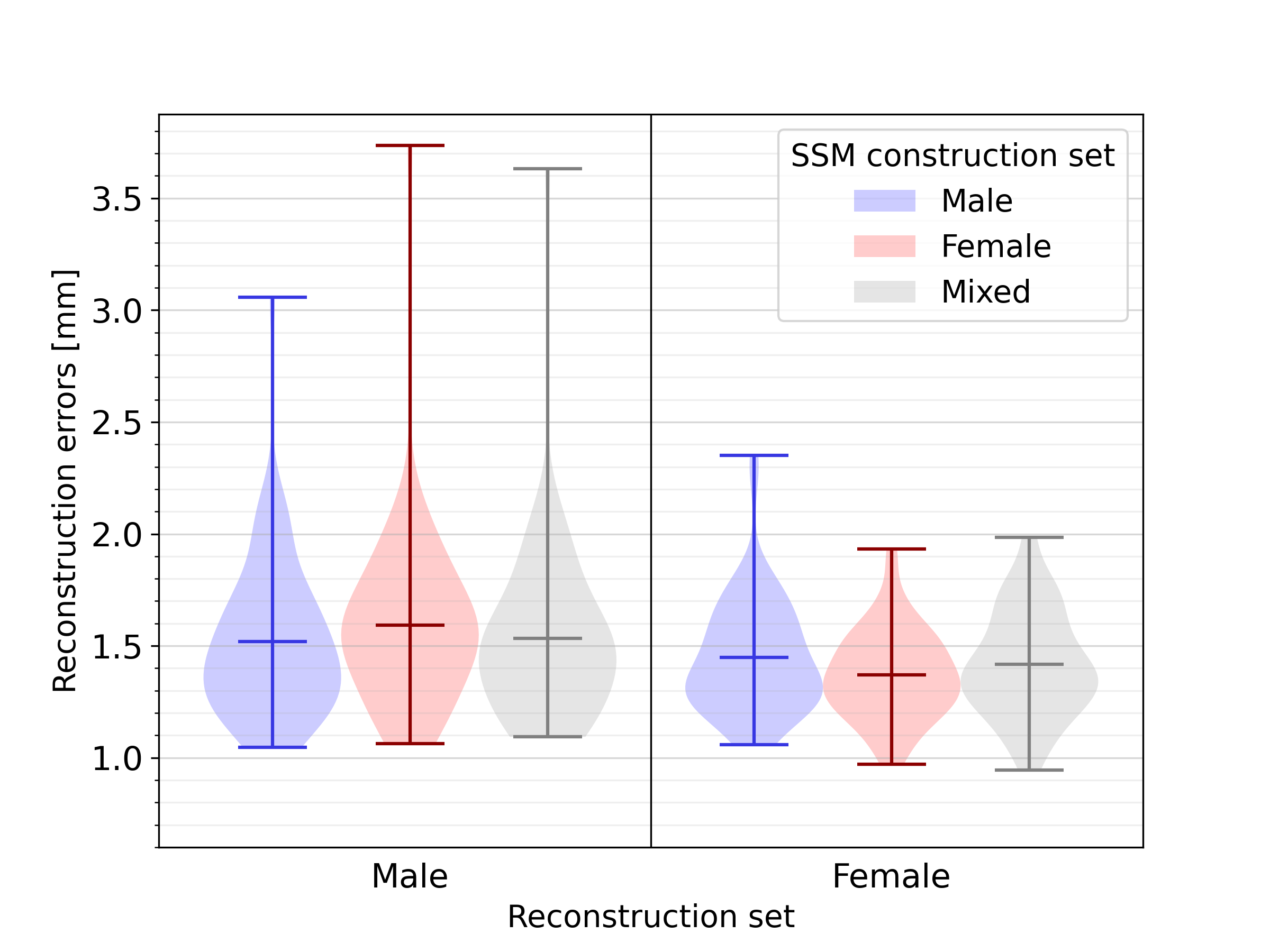}
	\caption{Heart size-independent reconstruction errors for male and female subjects in a series of SSMs constructed using a male, female or mixed dataset. The means, extrema and distributions are shown using violinplots.}
	\label{fig:malefemale_nosize}
\end{figure}

Figure~\ref{fig:malefemale_nosize} shows the effect of sex-related anatomical differences, excluding heart size, on representativeness of biventricular SSMs. Similarly to Figure~\ref{fig:malefemale}, the reconstruction of female hearts was associated with lower reconstruction errors than the reconstruction of male hearts, but the difference is lower when excluding heart size from the analysis. The interpretation of representativeness of opposite-, same- or mixed-sex SSMs remains the same as that of the SSMs including heart size, laid out in Section \ref{PCdistributions}.

\section{Impact of sex-related anatomical differences}
\label{app1}

\begin{figure}[!t]
	\centering
	\begin{subfigure}{0.38\textwidth}
		\centering
		\includegraphics[width=\textwidth]{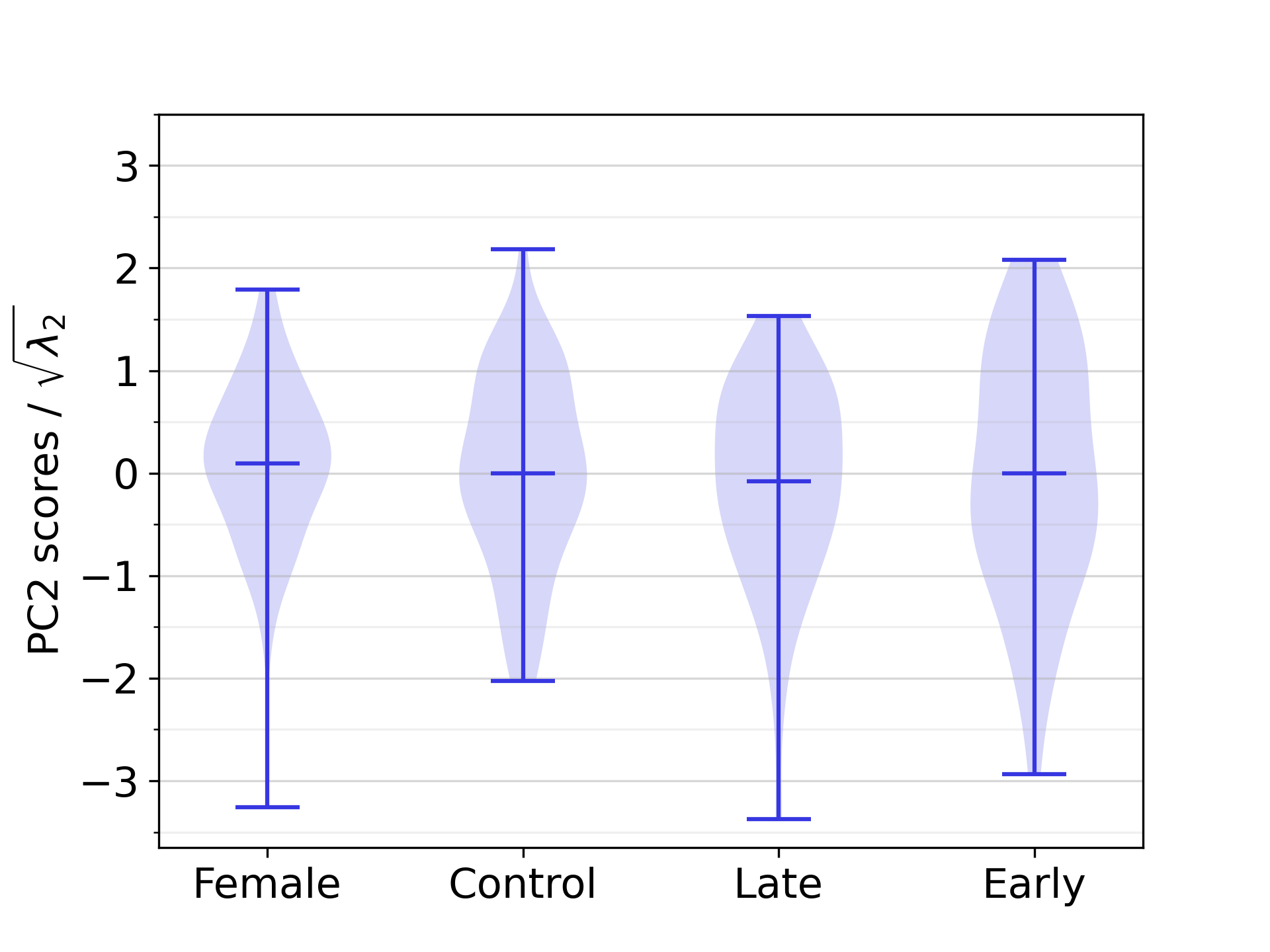}
		\caption{\footnotesize{Second principal component (PC2).}}
		\label{fig:PC2_subgroups}
	\end{subfigure}
	\hfill
	\begin{subfigure}{0.38\textwidth}
		\centering
		\includegraphics[width=\textwidth]{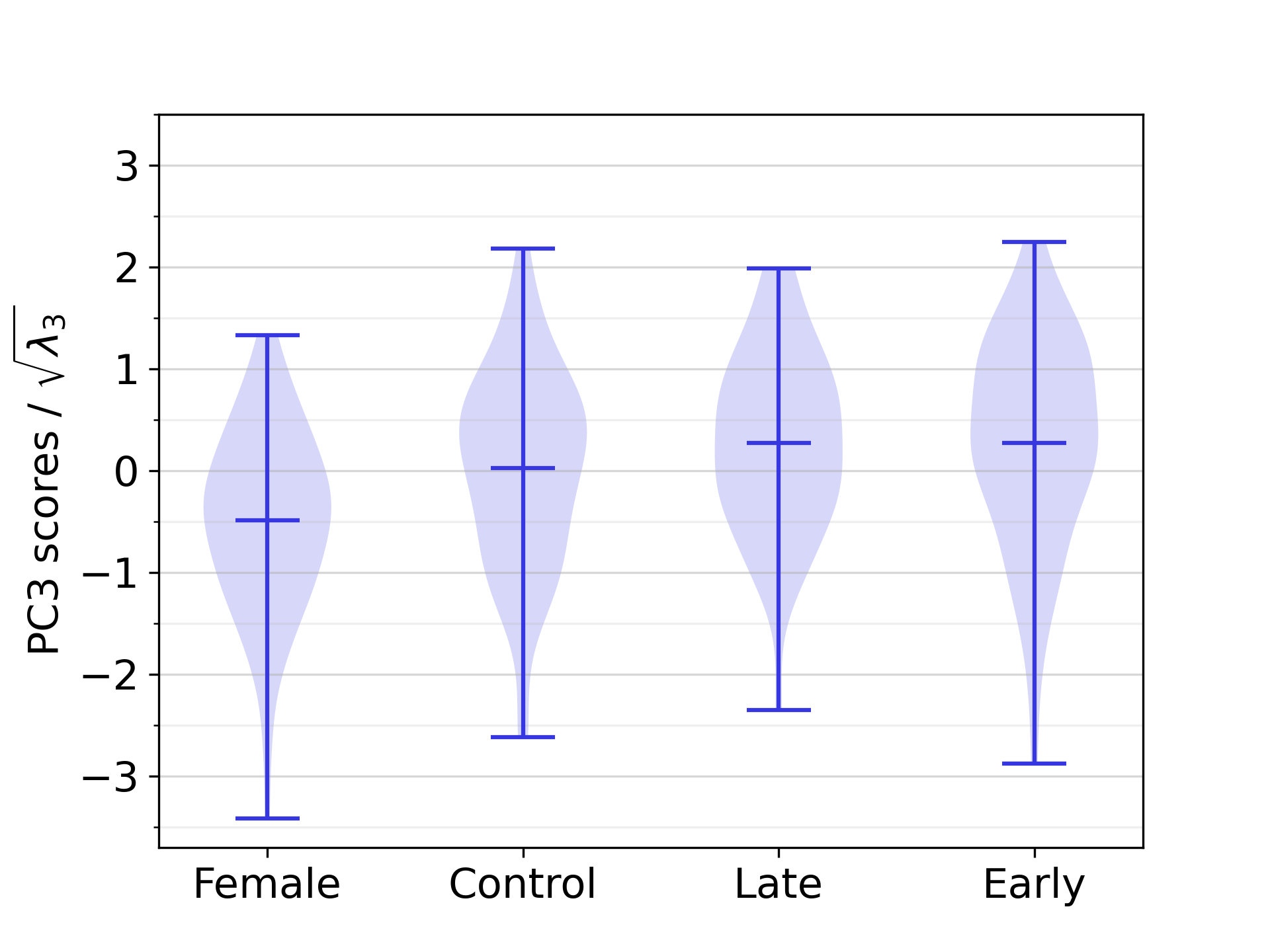}
		\caption{\footnotesize{Third principal component (PC3).}}
		\label{fig:PC3_subgroups}
	\end{subfigure}
	\caption{Distribution of scores for various PCs for the different demographic subgroups. The PC scores are scaled by their standard deviation $\sqrt{\lambda}$.}
	\label{fig:PC_subgroups}
\end{figure}

As described in Section \ref{PCdistributions}, statistical differences between the PC scores for distinct anatomical groups are only detected for PC1. This is displayed in Figure~\ref{fig:PC1_subgroups}. Distributions of PC2 and PC3 scores across the demographic groups are instead reported in Figure~\ref{fig:PC_subgroups}.

\section{Construction of synthetic cohort}
\label{app3}

\begin{figure*}
	\centering
	\includegraphics[width=0.98\textwidth]{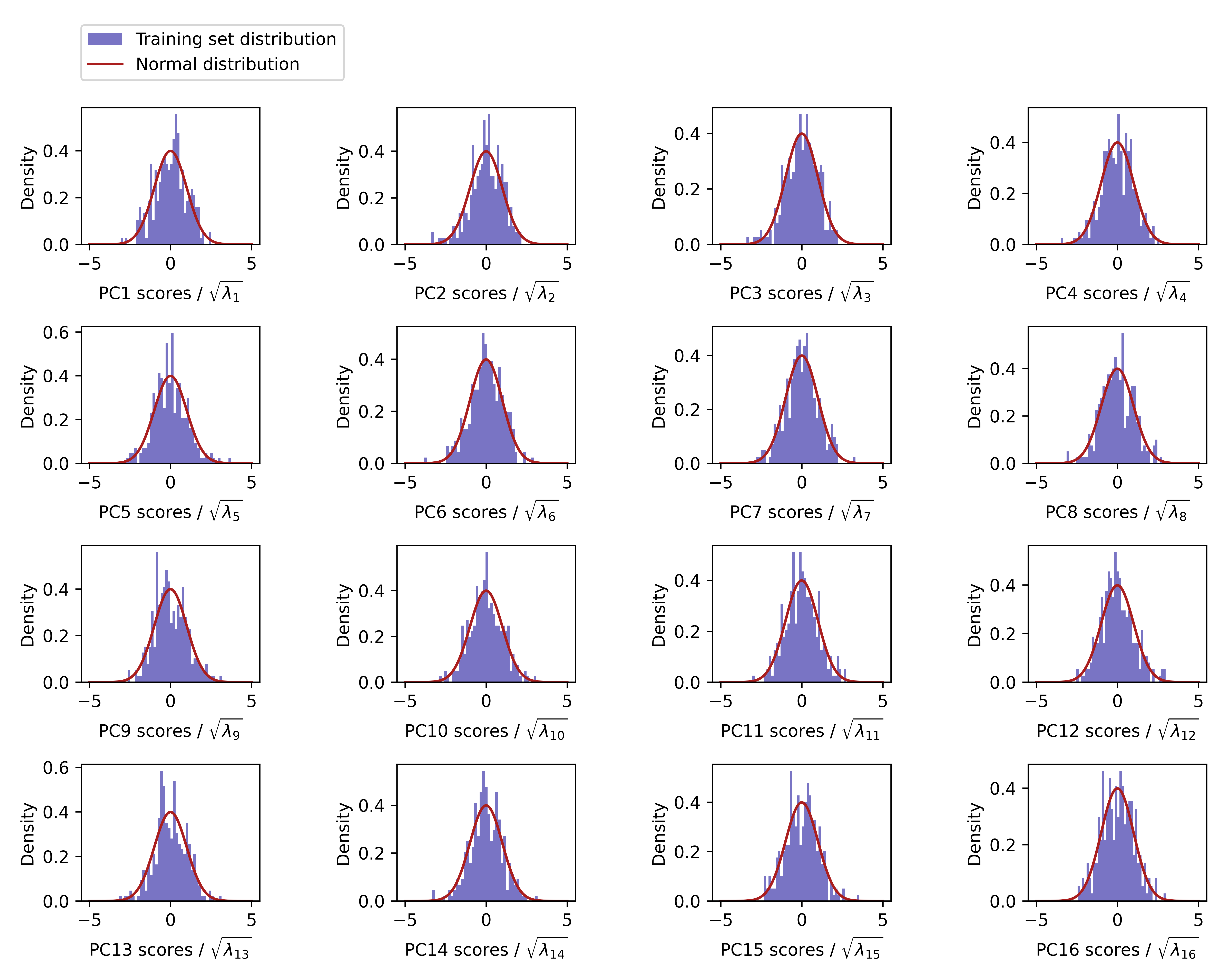}
	\caption{Distribution of the PC scores of the first 16 PCs in the full training set, compared to the normal distribution.}
	\label{fig:PC_distributions}
\end{figure*}

Figure~\ref{fig:PC_distributions} shows that the PC scores in the training set are approximately normally distributed. 

\newpage
\bibliographystyle{unsrtnat}
\bibliography{library_LVS}  

\end{document}